\documentclass[a4paper,onecolumn,11pt,accepted=2026-03-26]{quantumarticle}
\pdfoutput=1
\usepackage{geometry,amssymb,latexsym,amsmath,graphicx}
\usepackage[utf8]{inputenc}
\usepackage[normalem]{ulem}
\usepackage{tikz,ifthen}
\usepackage{tikzit}

\usepackage{thmtools} 
\usepackage{thm-restate}

\tikzstyle{triangle}=[style=isosceles triangle,isosceles triangle apex angle=60,draw,rotate=90, fill=black!00, minimum size =0.25cm,inner sep =0pt]
\tikzstyle{losange}=[style=diamond,draw, fill=black!0, minimum size =0.3cm,inner sep =0pt]
\tikzstyle{rond}=[style=circle,draw, fill=black!00, minimum size =0.25cm,inner sep =0pt]

\tikzstyle{gtriangle}=[style=isosceles triangle,isosceles triangle apex angle=60,draw=black, fill=black!30, rotate=90, minimum size =0.4cm,inner sep =0pt]
\tikzstyle{glosange}=[style=diamond,draw=black,  fill=black!30, minimum size =0.45cm,inner sep =0pt]
\tikzstyle{grond}=[style=circle,draw=black,  fill=black!30, minimum size =0.4cm,inner sep =0pt]
    
\tikzstyle{gn}=[fill=green, draw=black, shape=circle,  tikzit category=ZX, tikzit fill=green, tikzit draw=black, tikzit shape=circle, inner sep=0.1em]
\tikzstyle{rn}=[fill=red, draw=black, shape=circle, tikzit fill=red, tikzit draw=black, tikzit category=ZX, tikzit shape=circle, inner sep=0.1em]
\tikzstyle{divide}=[regular polygon, regular polygon sides=3, shape border rotate=90, draw=black,fill=white, inner sep=2pt, tikzit category=scal, rounded corners=0.8mm]
\tikzstyle{black}=[fill=black, draw=black, shape=circle, tikzit fill=black, tikzit draw=black, tikzit shape=circle, tikzit category=IH, inner sep=0.8pt]
\tikzstyle{gather}=[fill=white, draw=black, tikzit category=scal, rounded corners=0.8mm, regular polygon, regular polygon sides=3, shape border rotate=-90, inner sep=2pt]
\tikzstyle{ggen}=[fill=white, draw=black, shape=rectangle, rounded corners=2mm,  line width=1pt, tikzit draw=red, tikzit category=scal]
\tikzstyle{white}=[fill=white, draw=black, shape=circle, inner sep=2pt, tikzit category=IH]
\tikzstyle{mbox}=[fill=white, draw=black, rounded rectangle, rounded rectangle west arc=none, tikzit category=scal, tikzit shape=rectangle]
\tikzstyle{A}=[fill=white, shape=circle, tikzit category=scal, inner sep=1pt]
\tikzstyle{ggreen}=[fill=green, draw=black, shape=circle, tikzit category=SZX, tikzit fill=green, tikzit draw=black, line width=1pt, inner sep=0.1em]
\tikzstyle{gred}=[fill=red, draw=black, shape=circle, rounded corners=2mm,  tikzit category=SZX, inner sep=0.1em, tikzit fill=red, line width=1pt]
\tikzstyle{ghad}=[fill=yellow, draw=black, shape=rectangle, tikzit category=SZX, tikzit shape=rectangle, tikzit fill=yellow, inner sep=0.1em, line width=1pt]
\tikzstyle{boxm}=[fill=white, draw=black, rounded rectangle, tikzit category=scal, tikzit shape=rectangle, rounded rectangle east arc=none]
\tikzstyle{box}=[draw=grey, shape=rectangle]
\tikzstyle{had}=[fill=yellow, draw=black, shape=rectangle, tikzit category=ZX, tikzit fill=yellow, tikzit draw=black, inner sep=0.1em]
\tikzstyle{gwhite}=[fill=white, draw=black, shape=circle, tikzit fill=white, tikzit shape=circle, line width=1 pt, inner sep=2 pt, tikzit draw=red]
\tikzstyle{gblack}=[fill=black, draw=black, shape=circle, tikzit fill=black, tikzit shape=circle, line width=1 pt, inner sep=2 pt, tikzit draw=red]
\tikzstyle{antipode}=[fill=red, draw=black, shape=rectangle, tikzit fill=red, tikzit draw=black, tikzit shape=rectangle, inner sep=2pt]
\tikzstyle{rdelay}=[kite, draw, kite vertex angles=90 and 45,rotate=90,scale=0.6,fill=gray!60!white]
\tikzstyle{ldelay}=[kite, draw, kite vertex angles=90 and 45,rotate=-90,scale=0.6,fill=gray!60!white]
\tikzstyle{mongr}=[fill=green, draw=green, shape=circle, inner sep=2pt]
\tikzstyle{monbl}=[fill=blue, draw=blue, shape=circle, inner sep=2pt]
\tikzstyle{bg}=[inner sep=0.7mm, minimum width=0pt, minimum height=0pt, fill=green, draw=white, very thick, shape=circle]
\tikzstyle{br}=[inner sep=0.7mm, minimum width=0pt, minimum height=0pt, fill=red, draw=white, very thick, shape=circle]
\tikzstyle{rmat}=[draw, signal, fill=zx_grey, signal to=east, signal from=west, inner sep=1pt, minimum height=6pt]
\tikzstyle{lmat}=[draw, signal, fill=zx_grey, signal to=west, signal from=east, inner sep=1pt, minimum height=6pt]
\tikzstyle{umat}=[draw, signal, fill=zx_grey, signal to=north, signal from=south, inner sep=1pt, minimum width=6pt]
\tikzstyle{dmat}=[draw, signal, fill=zx_grey, signal to=south, signal from=north, inner sep=1pt, minimum width=6pt]
\tikzstyle{arrow}=[->]
\tikzstyle{very thick}=[-, line width=1pt, tikzit draw=red]
\tikzstyle{pointille}=[dashed, -]
\tikzstyle{memo}=[dashed, -, draw=gray!80!white]
\tikzstyle{red}=[-, draw=red]
\tikzstyle{blue}=[-, draw=blue]
\tikzstyle{green}=[-, draw=green]
\tikzstyle{yellow}=[-, draw=orange]
\tikzstyle{arrow}=[->]
\tikzstyle{strike}=[-, tikzit draw={rgb,255: red,191; green,0; blue,64}, strike through]
\tikzstyle{strike'}=[-, tikzit draw=cyan, strike bend]

\tikzstyle{box2}=[shape=rectangle, text height=6ex, text depth=0.25ex, yshift=0mm, fill=blue, draw=black, minimum height=1mm, minimum width=5mm, font={\small}]
\tikzstyle{box}=[shape=rectangle, text height=1.5ex, text depth=0.25ex, yshift=0.5mm, fill=white, draw=black, minimum height=5mm, yshift=-0.5mm, minimum width=5mm, font={\small}]
\tikzstyle{Z dot}=[inner sep=0mm, minimum size=2mm, shape=circle, draw=black, fill={rgb,255: red,160; green,255; blue,160}]
\tikzstyle{gdot}=[minimum size=3mm, font={\scriptsize\boldmath}, shape=rectangle, rounded corners=1.3mm, inner sep=1mm, outer sep=-1.8mm, scale=0.8, tikzit shape=circle, draw=black, fill=green, tikzit draw=blue]
\tikzstyle{X dot}=[Z dot, shape=circle, draw=black, fill={rgb,255: red,220; green,0; blue,0}]
\tikzstyle{rdot}=[minimum size=3mm, font={\scriptsize\boldmath}, shape=rectangle, rounded corners=1.3mm, inner sep=1mm, outer sep=-1.8mm, scale=0.8, tikzit shape=circle, draw=black, fill=red, tikzit draw=blue]
\tikzstyle{grdot}=[minimum size=3mm, font={\scriptsize\boldmath}, shape=rectangle, rounded corners=1.3mm, inner sep=1mm,
 line width=1pt, outer sep=-1.5mm, scale=0.8, tikzit shape=circle, draw=black, fill=red, tikzit draw=blue]
 \tikzstyle{ggdot}=[minimum size=3mm, font={\scriptsize\boldmath}, shape=rectangle,  line width=1pt, rounded corners=1.3mm, inner sep=1mm, outer sep=-1.5mm, scale=0.8, tikzit shape=circle, draw=black, fill=green, tikzit draw=blue]

\usetikzlibrary{shapes.geometric}

\usetikzlibrary{calc}

\usepackage{hyperref}
\usepackage[backend=biber, doi=true]{biblatex}
\usepackage{mathabx}
\usepackage{stmaryrd}
\newtheorem{theorem}{Theorem}
\newtheorem{thm}{Theorem}

\newtheorem{lemma}{Lemma}
\newtheorem{fact}{Fact}

\newtheorem{corollary}{Corollary}

\newtheorem{proposition}{Proposition}

\newtheorem{definition}{Definition}
\newtheorem{remark}{Remark}

\newcommand{\qed}{\hspace{\stretch{1}} $\Box$}
\newenvironment{proof*}[1]{\noindent{\bf #1}}{\qed}
\newenvironment{proof}{\noindent {\bf Proof:}}{\qed}

\newcommand{\odd}[1]{\textup{\textsf{Odd}}({#1})}

\newcommand{\ket}[1]{\left| #1 \right\rangle}
\newcommand{\bra}[1]{\left\langle #1 \right|}

\newcommand{\E}{\textup {\textsf E}}

\newcommand{\N}{\textup {\textsf N}}
\newcommand{\X}{\textup {\textsf X}}
\newcommand{\Y}{\textup {\textsf Y}}
\newcommand{\Z}{\textup {\textsf Z}}
\newcommand{\M}{\textup {\textsf M}}
\newcommand{\s}{\textup {\textsf s}}

\newcommand{\cor}{\textup {cor}}

\newcommand{\infl}{\textup {imp}}
\newcommand{\ana}{\textup {ana}}

\newcommand{\interp}[1]{\left\llbracket #1 \right\rrbracket}

\title{Shadow Pauli Flow: Characterising Determinism in MBQCs involving Pauli Measurements}

\author[1]{Mehdi Mhalla}
\author[2]{Simon Perdrix}
\author[2,3]{Luc Sanselme}
\affil[1]{Universit\'e Grenoble Alpes, CNRS, Grenoble INP, LIG, F-38000 Grenoble, France}
\affil[2]{Inria Mocqua, LORIA, CNRS, Universit\'e de Lorraine,  F-54000 Nancy, France}
\affil[3]{Lycée Poincaré}

\date {}
\begin{document}

\maketitle

\begin{abstract}
We introduce a new characterisation of determinism in Measurement-Based Quantum Computing (MBQC).
The one-way model of computation consists in performing local measurements over a large entangled
state represented by a graph. The ability to perform an overall deterministic computation requires a
correction strategy because of the non-determinism of each measurement. The existence of such
correction strategy depends on the underlying open graph, which is a description of the resource
state together with the basis of the performed measurements. GFlow is a well-known graphical
characterisation of robust determinism in MBQC when every measurement is performed in some specific
planes of the Bloch sphere. While Pauli measurements are ubiquitous in MBQC, GFlow fails to be
necessary for determinism when a measurement-based quantum computation involves Pauli measurements.
As a consequence, Pauli Flow  was designed as a generalisation of GFlow to handle MBQC with Pauli
measurements: Pauli Flow guarantees robust determinism, however it has been shown more recently that
it fails to be a necessary condition.

 Our contribution is twofold. First, we demonstrate that Pauli Flow is actually necessary for robust
 determinism in a weaker sense: given an open graph, i.e.~a resource state, a deterministic
 computation can be driven if only if it has a Pauli Flow. However, the Pauli Flows do not reflect
 all the possible correction strategies over a particular resource state,  and properties like
 measurement order or computational depth are not necessarily reflected by a Pauli Flow. Thus, to
 characterise determinism in full generality, we introduce  a further extension called \emph{Shadow
 Pauli Flow} that  we prove necessary and sufficient for robust determinism: An MBQC is robustly
 deterministic if and only if its correction strategy is consistent with a Shadow Pauli Flow.
 Furthermore we show that Shadow Pauli Flow can be computed in polynomial time.

\end{abstract}

\section{Introduction}

Measurement-Based Quantum Computing (MBQC), also called one-way model, has been introduced by Briegel and Raussendorf in~\cite{raussendorf2001one,gross2007measurement,raussendorf2009measurement,Briegel:2009jk}. The key idea is to perform single-qubit measurements over a large resource (entangled) state.
It had shown very interesting properties like robustness to noise~\cite{raussendorf2007topological} and is also useful for photonic quantum computing proposals~\cite{bartolucci2021fusion} via a variant called FBQC. Another application is to decrease the quantum depth: for instance the Quantum Fourier Transform can be approximated in constant quantum depth using MBQC~\cite{browne2010computational} and thus provides a logarithmic speed-up compared to quantum circuits.

A fundamental property of MBQC is its probabilistic behavior: every measurement leads to two
possible outcomes. In this context, performing an overall ``deterministic" computation -- e.g. for
implementing a unitary evolution -- requires a correction strategy, an adaptive computation which
depends on the classical outcomes of the measured qubits, so that the output of the overall
computation does not depend on the intermediate measurement outcomes. Robust determinism strengthens
this requirement by demanding that determinism holds at every stage of the computation, with all
measurement outcomes occuring with the same probability and remaining deterministic for any variation
of the planes measurements angles.

Robust determinism is in particular well-suited for variational quantum computing. Several recent
works \cite{ferguson2021measurement,schroeder2023deterministic} points out the necessity of
determinism in Measure\-ment-based Variational Quantum Eigensolver. Measurement-based Variational Quantum Eigensolver is particularly promising in terms of physical implementation.
Robust determinism guarantees the so-called \emph{deterministic Ansätze}, indeed determinism is preserved when angles, i.e. the parameters, vary. 

The existence of  a correction strategy essentially depends on two parameters: the initial resource state
and the performed measurements. In MBQC, every measurement can be performed in three possible planes of the Bloch sphere (the so-called $XY$, $YZ$, and $XZ$-planes). The resource state and the measurements can be abstracted away into an \emph{open graph},  i.e., a graph that describes the resource state together with a labelling of the vertices with the corresponding measurement plane (e.g. $XZ$). 
In~\cite{browne2007generalized}, it has been shown that a robust deterministic computation is
possible if and only if the corresponding open graph has a GFlow. Hence GFlow, an easy to compute
graphical property, characterises robust determinism in MBQC\@. GFlow had several applications in MBQC
but also in related topics like ZX-calculus~\cite{duncan2010rewriting} and quantum circuit
optimisation~\cite{duncan2020graph,backens2021there}.

Pauli measurements are ubiquitous in MBQC, for instance in the context of error correcting codes, or
for implementing Clifford transformations. Some graphical transformations can be performed using
Pauli measurements so that they can be used to locally transform a resource state into another one
satisfying for instance some architecture constraints. More generally, it is reasonable to assume
that a significant portion of the measurements of a practical MBQC are Pauli measurements. This is for instance the case when using as a resource a constraint initial state like a cluster state or a triangular grid \cite{DBLP:journals/ijuc/MhallaP13}.  In terms
of determinism, Pauli measurements satisfy some particular properties that are not captured by
GFlow. For instance, an MBQC involving an X measurement can be deterministic, even though the corresponding open graph has no GFlow, regardless of whether the X measurement is considered an XY or XZ measurement. 
Hence a generalisation of GFlow, called Pauli Flow has been introduced in the original paper
introducing GFlow~\cite{browne2007generalized}. Pauli Flow has been proved to be a sufficient
condition for robust determinism. Pauli Flow has been proved to be also a necessary condition for
determinism when all measurements are real\footnote{i.e.~measurements described by real observables:
$X$, $Z$ and any measurement in the $XZ$-plane.}, but is not necessary in
general~\cite{perdrix2017determinism}. Notice that recently, Pauli Flow has been used to obtain very
efficient circuit optimisations~\cite{simmons2021relating}  and  rewrite rules for MBQC patterns
preserving Pauli Flow has been  derived in~\cite{mcelvanney2022complete}.

We introduce an extension of the Pauli Flow called \emph{Shadow Pauli Flow} that we
prove to be a necessary and sufficient condition for robust determinism:
{\begin{thm}\!\!\!\ref{thm:CNS}  {\bf (Informal)}
 An MBQC is robustly deterministic if and only if it is consistent with a Shadow Pauli Flow on its open graph. 
\end{thm}}

To prove this theorem,  we first show that when all
Pauli measurements are performed at the beginning of the computation,  the Pauli  Flow is
actually necessary and sufficient for robust determinism. The rest of the proof is obtained  by
showing how Pauli measurements can be moved forwards or backwards in an MBQC, and how such moves
preserve robust determinism and/or Shadow Pauli Flow existence.
 
 {Notice that it implies that Pauli Flow actually also characterises robust determinism but in a weaker way:
\begin{thm}\!\!\!\ref{thm:PFCNS}  {\bf (Informal)}
A resource state, described as an open graph, can be used to perform a robustly deterministic MBQC if and only if it has a Pauli Flow. 
 \end{thm}
 
 The first theorem is stronger as, in particular, it relates the order in which the measurements are performed -- and hence the depth of the MBQC -- with  the order of the corresponding flow. 
 }
 
It is common to discard the Pauli measurements in an analysis of MBQC considering that they can be moved first \cite{raussendorf2009measurement}. We strengthen this observation showing that moving the Pauli measurements first can be done in a way that preserves robust determinism. Furthermore, the robustness property applies for the original pattern of measurements which can be relevant in practice as it may be impossible to move some Pauli measurements in some implementations as one can reuse some qubits or have restrictions on the position of the qubits measured at some time step.

We show that given a partial order on measured qubits and an open graph, a Shadow Pauli Flow can be computed in polynomial time. 

The paper is structured as follows. We start by recalling the basic definitions of MBQC patterns and
their semantics in section~\ref{sec:defMBQC}, then in section~\ref{sec:defExtended Pauli Flow} we
define the Shadow Pauli Flow.  In section~\ref{sec:CS}, we show that Shadow Pauli Flow is sufficient
for robust determinism. We demonstrate in Section~\ref{sec:NecGFlow} that GFlow is necessary when
there are no Pauli measurements (fixing a  mistake in the proof published
in~\cite{browne2007generalized}), then we show in section~\ref{sec:NecPauliFirst} that Pauli Flow is
necessary when the Pauli measurements are done first, and  that Shadow Pauli Flow is sufficient for
a general pattern in section~\ref{sec:NecExtended Pauli Flow}. Finally, in section~\ref{sec:algo},
we introduce an efficient algorithm for Shadow Pauli Flow.

\section{Determinism in MBQC}

A {resource state for} MBQC can be informally represented as follows: 

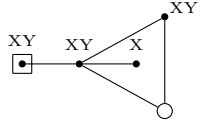
\[\begin{tikzpicture}
	\begin{pgfonlayer}{nodelayer}
		\node [style=black] (8) at (-2, 0) {};
		\node [style=black] (10) at (-0.5, 0) {};
		\node [style=black] (11) at (1.75, 1.25) {};
		\node [style=black] (12) at (1, 0) {};
		\node [style=none] (13) at (-2, 0.625) {\tiny XY};
		\node [style=none] (14) at (-0.5, 0.5) {\tiny XY};
		\node [style=none] (15) at (2.25, 1.5) {\tiny XY};
		\node [style=none] (17) at (1, 0.5) {\tiny X};
		\node [style=white] (18) at (1.75, -1.25) {};
		\node [style=none] (23) at (-2.25, 0.25) {};
		\node [style=none] (24) at (-1.75, 0.25) {};
		\node [style=none] (25) at (-1.75, -0.25) {};
		\node [style=none] (26) at (-2.25, -0.25) {};
	\end{pgfonlayer}
	\begin{pgfonlayer}{edgelayer}
		\draw (8) to (10);
		\draw (10) to (12);
		\draw (10) to (11);
		\draw (11) to (18);
		\draw (10) to (18);
		\draw (23.center) to (24.center);
		\draw (24.center) to (25.center);
		\draw (25.center) to (26.center);
		\draw (26.center) to (23.center);
	\end{pgfonlayer}
\end{tikzpicture}
\]
where the underlying graph represents the entangled resource state of the computation, the outputs
qubits are represented by white vertices and the input qubits are surrounded by a square. For each measured qubit, a description of the measurement basis is given: it is either a Pauli measurement ($X$, $Y$, or $Z$) or a measurement in a plane of the Bloch sphere ($XY$, $XZ$, or $YZ$). Notice that such a graphical representation  provides a partial information of the MBQC: the actual angles describing the measurement basis, the order of the measurements and the correction strategy are not depicted. In this section, we provide a more formal description of MBQC based on the Measurement calculus~\cite{danos2007measurement}. 

\label{sec:mbqc}
\subsection{MBQC syntax and semantics}\label{sec:defMBQC}
We use the notation $\prec$ to denote a strict partial order, i.e.~a transitive asymmetric binary
relation. The reflexive closure of  $\prec$ is denoted $\preccurlyeq$. Conversely, given a
non strict partial order $\preccurlyeq$, we denote $\prec$ its irreflexive kernel.
We use $\le$ (resp.~$<$) to denote a (resp.~strict) total order.

The measurement calculus is a formal framework for describing MBQC using patterns, a pattern being a sequence of commands. 
There are five kinds of commands: $\N,\E,\M, \X,\Z$; describing how the qubits are initialized ($\N$), entangled ($\E$),  measured ($\M$), and corrected ($\X, \Z$). 

A pattern is made of two parts: a description of the initial entanglement and then a description of the measurements and their associated corrections. Given a graph $G$ whose vertices $V(G)$ represent the qubits -- including a set $I$ of input qubits -- the commands $\N_{V(G)\setminus I}$ and $\E_G$ describe the initial entanglement of the computation.  

The measurement of a qubit  $u$ is denoted by  $\M_u^{\lambda,\alpha}$, where $(\lambda,\alpha)$
characterizes the measurement basis. $\lambda$ is either a Pauli observable or a plane defined by
two Pauli observables\footnote{$\lambda$ can be seen as a set of one or two Pauli observables
\emph{e.g.} $\{\X\}$ for $X$-measurement and $\{\X,\Y\}$ for $XY$-plane measurement.}, and $\alpha$
is the measurement angle. Notice that the measurements of the input qubits are restricted to the $XY$-plane of the Bloch sphere.

The necessary corrections (Pauli operators $\X_A^{s_u},\Z_A^{s_u}$ applied to a subset of vertices $A$ depending on the classical  outcome $s_u$ of the measurement of $u$) are also represented in a pattern. Each qubit is measured at most once, and the unmeasured ones form the set $O$ of output qubits. 

\begin{definition}[Pattern]\label{def:pattern}

	A pattern $\mathcal P:I\to O$, with $I$ and $O$ two finite sets,  is inductively defined as: 
	\begin{itemize}
		\item for any simple undirected finite graph  $G$, and any $I\subseteq V(G)$,  
		\[\E_G \N_{V(G)\setminus I} :I\to V(G)\] is a pattern;
	\item for any pattern $\mathcal P:I\to O$, any qubit $u\in O$,  any subsets $A,B\subseteq O\setminus \{u\}$, any $\lambda\subsetneq \{\X,\Y,\Z\}$ s.t. $ |\lambda|\ge 1$ and any angle $\alpha \in [0,2\pi)$ 
	\[\X^{\s_u}_A\Z^{\s_u}_B\M_u^{\lambda, \alpha}\mathcal P: I \to O\setminus \{u\}\] is a pattern where $\alpha\in \{0,\pi\}$ when $|\lambda|=1$, and $\lambda \subseteq \{\X,\Y\}$ when $u\in I$.

	\end{itemize}
\end{definition}

\begin{remark}

	Some more general form of patterns have been defined in the literature~\cite{danos2007measurement}. It has been
	proved~\cite{danos2007measurement} however that, using a standardisation procedure, any pattern can be transformed,
	preserving the semantics, into a pattern of the form of Definition~\ref{def:pattern}.

\end{remark}

\begin{remark}

	Notice that Pauli corrections can also be integrated to the upcoming measurements, leading to
	adaptive measurements instead of Pauli corrections. The two presentations being equivalent
	\cite{danos2007measurement} we choose w.l.o.g.~to represent MBQC patterns with Pauli corrections.

\end{remark}

We use the following notation to denote an arbitrary pattern:
\[\left(\prod_{u\in O^c}^< \X_{\mathbf x(u)}^{\s_u}\Z_{\mathbf z(u)}^{\s_u}\M_u^{\lambda_u,
\alpha_u}\right)\E_G\N_{I^c}= \X_{\mathbf x(u_k)}^{\s_{u_k}}\Z_{\mathbf
z(u_k)}^{\s_{u_k}}\M_{u_k}^{\lambda_{u_k}, \alpha_{u_k}}\ldots \X_{\mathbf
x(u_1)}^{\s_{u_1}}\Z_{\mathbf z(u_1)}^{\s_{u_1}}\M_{u_1}^{\lambda_{u_1}, \alpha_{u_1}}\E_G\N_{I^c}
\]
where $<$ is the total order over $O^c:=V(G)\setminus O $ s.t. $u_1<\ldots <u_k$.

Each MBQC pattern has an underlying structure that is an extension of graph states taking into account inputs, outputs and measurements that is called an open graph.

\begin{definition}[Open graph] The
	quadruplet $(G,I,O,\lambda)$ is the underlying open graph of a pattern $\left(\prod\limits_{u\in O^c}^< \X_{\mathbf
	x(u)}^{\s_u}\Z_{\mathbf z(u)}^{\s_u}\M_u^{\lambda_u, \alpha_u}\right)\E_G\N_{I^c}$ with $\lambda: u\mapsto \lambda_u$.

\end{definition}

An open graph is an abstraction of a pattern where three properties are essentially abstracted away: the order (or scheduling) of the measurements, the actual angle of the measurements (only the measurement-plane is kept for non-Pauli measurements), and finally the corrections.

Several patterns can lead to the same open graph, it is however convenient to consider, among all these patterns, those which share the same correction strategy (represented as a function from non output qubits to sets of non input qubits) and have compatible measurement schedulings (represented as a partial order):

\sloppy
\begin{definition}[Consistent Pattern]

	Given $p:O^c \to 2^{I^c}$ and $\prec$, a partial order, a pattern $\left(\prod_{u\in O^c}^< \X_{\mathbf x(u)}^{\s_u}\Z_{\mathbf
	z(u)}^{\s_u}\M_u^{\lambda_u, \alpha_u}\right)\E_G\N_{I^c}$ is said to be \emph{consistent} with 
	 $(p,\prec)$  if $\forall u\in O^c$, $\mathbf
	x(u) = \{v\in p(u)~|~u\prec v\}$, $\mathbf z(u) = \{v\in \odd{p(u)}~|~u\prec v\}$, and $\forall
	v\in O^c$, $u\prec v \Rightarrow u<v$,
{where $\odd{p(u)}$ is the set of vertices that have an odd number of neighbours in $p(u)$.}

\end{definition}

A pattern involves quantum measurements: each non-output qubit $u$ is measured, which produces a
classical outcome $m(u)\in \{0,1\}$.

$\M^{\lambda, \alpha}$ is a measurement in the basis $\{\ket {+_{\alpha}^{\lambda}},
\ket{-_{\alpha}^{\lambda}}\}$ defined as follows:
\begin{itemize}
	\item in case of a Pauli measurement i.e.~when $|\lambda|=1$,

\begin{equation}
\label{measurementBasis}
\hspace{-1cm}\ket {+_0^\lambda}=\begin{cases}
  \frac 1  {\sqrt{2}}  (\ket 0 + \ket 1)  &~\text{if}~\lambda = \{\X\}
\\  \frac 1  {\sqrt{2}} (\ket 0 + i \ket 1)  &~\text{if}~\lambda = \{\Y\}
\\    \ket 0   &~\text{if}~\lambda = \{\Z\}
\end{cases}~\text{and}~\ket {-_0^\lambda}= \begin{cases}
  \frac 1  {\sqrt{2}}  (\ket 0 - \ket 1)  &~\text{if}~\lambda = \{\X\}
\\  \frac 1  {\sqrt{2}} (\ket 0 - i \ket 1)  &~\text{if}~\lambda = \{\Y\}
\\    \ket 1   &~\text{if}~\lambda = \{\Z\}
\end{cases}
\end{equation}
and $\ket {+_{\pi}^\lambda}=\ket {-_0^\lambda}$ and $\ket {-_{\pi}^\lambda}=\ket {+_0^\lambda}$,
	\item and when $|\lambda|=2$, 
   $\ket {+_\alpha^\lambda}=  \frac 1  {\sqrt{2}}  (\ket {+_0^P} + e^{i\alpha}\ket {-_0^P})$  and
   $\ket {-_\alpha^\lambda}=  \frac 1  {\sqrt{2}}  (\ket {+_0^P} - e^{i\alpha}\ket {-_0^P})$ where
	 $P\notin \lambda$.
\end{itemize}

We also define $P(\alpha)$ as the unitary such that $P(\alpha) \ket {+^P_0}= \ket {+^P_0}$ and $P(\alpha) \ket {-^P_0}= e^{-i\alpha} \ket {-^P_0}$.

The action of a pattern on a quantum input state  can then be
described as a collection of linear maps, one for each possible branch of the computation which corresponds to a sequence of classical outcomes:

\begin{definition}[Branch semantics]

	Given a pattern $\mathcal P:I\to O$ and  a sequence of classical outcomes $m:O^c\to \{0,1\}$, let
	$\interp {\mathcal P}_m: \mathbb C^{2^I}\to \mathbb C^{2^O}$ be inductively defined as
	\[\interp{\E_G \N_{V(G)\setminus I}}_m = \ket \phi \mapsto   E_G \ket +_{I^c} \ket{\phi}_I =  \prod_{(u,v)\in G}\Lambda Z_{u,v}\left(\bigotimes_{u\in V(G)\setminus I}\frac {\ket {0_u}+\ket{1_u}}{\sqrt 2}\right) \otimes \ket \phi\]
\[ \interp{\X^{\s_u}_A\Z^{\s_u}_B\M_u^{\lambda, \alpha}\mathcal P}_m =
	\begin{cases}
		\bra{+^{\lambda_u}_{\alpha_u}}_u \interp{\mathcal P}_{m'}, &~\text{if}~m(u) = 0\\\
		X_AZ_B\bra{-^{\lambda_u}_{\alpha_u}}_u \interp{\mathcal P}_{m'}, &~\text{if}~m(u) = 1
	\end{cases}
\]
where $m'$ is the restriction of $m$ to $O^c\setminus  \{u\}$,
$\{\ket {+_{\alpha_u}^{\lambda_u}}, \ket {-_{\alpha_u}^{\lambda_u}}\}$ is the measurement basis
defined in previous definition, and $\Lambda Z$ is the control-$Z$ operator.

\end{definition}

The overall semantics of a pattern can then be described as a superoperator:

\begin{definition}[Semantics]

	Given a pattern $\mathcal P:I\to O$, its semantics is the superoperator $\interp {\mathcal P} =
	\rho \mapsto \sum_{m\in \{0,1\}^{O^c}} \interp {\mathcal P}_m \rho  \interp {\mathcal P}_m^\dagger$.

\end{definition}

A pattern is deterministic if the overall evolution does not depend on the intermediate classical
outcomes. We consider a robust version of determinism which is  \emph{strong} (all branches occur
with the same probability), \emph{uniform} (for any variation of the measurement angles the pattern
remains  strongly deterministic), and \emph{stepwise} (every partial computation is also uniform and
strongly deterministic).

\begin{definition}[Robust determinism]

 We inductively define {robustly deterministic} patterns as:
\begin{itemize}
\item $\E_G \N_{V(G)\setminus I}$ is robustly deterministic;
\item $\X^{\s_u}_A\Z^{\s_u}_B\M_u^{\lambda, \alpha}\mathcal P$ with $|\lambda| = 1$ is robustly deterministic when $ \mathcal P$ is robustly deterministic and $\forall \rho$, $\bra{+^{\lambda}_{\alpha}}_u \interp{\mathcal P}(\rho)\ket{+^{\lambda}_{\alpha}}_u = \bra{-^{\lambda}_{\alpha}}_u Z_BX_A\interp{\mathcal P}(\rho)X_AZ_B\ket{-^{\lambda}_{\alpha}}_u$    

\item $\X^{\s_u}_A\Z^{\s_u}_B\M_u^{\lambda, \alpha}\mathcal P$ with $|\lambda| = 2$  is robustly deterministic when $ \mathcal P$ is robustly deterministic and $\forall \epsilon \in \mathbb R$,  $\forall \rho$, $\bra{+^{\lambda}_{\alpha+\epsilon}}_u \interp{\mathcal P}(\rho)\ket{+^{\lambda}_{\alpha+\epsilon}}_u = \bra{-^{\lambda}_{\alpha+\epsilon}}_u Z_BX_A\interp{\mathcal P}(\rho)X_AZ_B\ket{-^{\lambda}_{\alpha+\epsilon}}_u$   

\end{itemize}

\end{definition}

\subsection{Characterising Determinism}\label{sec:defExtended Pauli Flow}

A central question in MBQC is to decide whether a deterministic computation can be driven on a
given resource, represented as an open graph $(G,I,O,\lambda)$. Several sufficient conditions  for
robust determinism have been introduced: Causal Flow~\cite{danos2006determinism} which has been generalized to GFlow and
Pauli Flow~\cite{browne2007generalized}. GFlow has been proved to be a necessary condition for robust
determinism when all measurements are performed in a plane ($\forall u \in O^c, |\lambda_u|=2$), and
Pauli Flow has been proved to be  necessary when all measurements are real ($\forall u \in O^c,
\lambda_u\subseteq \{\X,\Z\}$),  whereas counter examples are known in the general case.

The correction strategy in MBQC fundamentally relies on some fixed-point properties of graph states:
for any subset $D$ of (non input) qubits, the Pauli operator $X_D Z_{\odd{D}}$ 
 is called a stabilizer, because it leaves the resource state invariant.  The local operation, on every qubit $v\in V$, of this stabilizer is called the \emph{action of $D$ on $v$}, denoted $Act^D_v$ and  thus defined as:

 \[Act^D_v=\begin{cases} X \text{ if } v\in D\setminus \odd D 
\\ Z \text{ if } v\in \odd D\setminus D
\\ Y  \text{ if } v\in D\cap \odd D
  \\I \text{ Otherwise.} \end{cases}\]
The action on a vertex can be used to actually correct a measurement: when the action of $D$ on a vertex $u$ anti-commute with each element of $\lambda_u$ according to which $u$ is measured, the action of $D$ turns one measurement projector into the other, performing the desired correction. Hence,  we denote  
 \[ \cor(D)=\{u,  \forall P \in \lambda_u, [Act^D_u ,P]\neq 0 \}\] 

GFlow is based on this idea and consists in finding for every measured qubit $u$ a set $D$ of correctors such that (i) $u\in \cor(D)$ and (ii) $D$ acts trivially on all qubits measured before $u$ to avoid uncontrolled side effects.

The extension from GFlow to Pauli Flow is based on the following additional property: a correction can act on an already measured qubit as long as this qubit has been measured in the appropriate Pauli basis, in this case the qubit is not impacted by the action of a corrector. Hence,  
we say that the action of $D$ \emph{impacts} a vertex $v$  if the action of $D$ on $v$ anticommutes with one Pauli element of $\lambda_v$ and denote  
\[\infl(D)=\{v, \exists P\in \lambda_v, [Act^D_v ,P]\neq0 \}\]
\begin{figure}
\begin{tikzpicture}
	\begin{pgfonlayer}{nodelayer}
		\node [style=losange] (16) at (0, 0) {};
		\node [style=black] (0) at (0, 0) {};
		\node [style=none] (1) at (0, -0.5) {\tiny XY};
		\node [style=none] (2) at (-1, 2) {};
		\node [style=none] (3) at (-1, -2) {};
		\node [style=none] (4) at (1, 2) {};
		\node [style=none] (5) at (1, -2) {};
		\node [style=black] (6) at (-2, 1) {};
		\node [style=black] (7) at (-2, -1.25) {};
		\node [style=black] (8) at (-2.75, 0) {};
		\node [style=black] (9) at (2, 1.25) {};
		\node [style=rond] (16) at (2, -1.25) {};
		\node [style=black] (10) at (2, -1.25) {};
		\node [style=rond] (11) at (3.25, 0) {};
		\node [style=black] (11) at (3.25, 0) {};
		\node [style=black] (12) at (3, -2) {};
		\node [style=none] (13) at (2, 0.75) {\tiny XY};
		\node [style=none] (14) at (2, -1.75) {\tiny XZ};
		\node [style=none] (15) at (3.25, -0.5) {\tiny XY};
	\end{pgfonlayer}
	\begin{pgfonlayer}{edgelayer}
		\draw [style=pointille] (2.center) to (3.center);
		\draw [style=pointille] (4.center) to (5.center);
		\draw (7) to (10);
		\draw (0) to (11);
		\draw (11) to (7);
	\end{pgfonlayer}
\end{tikzpicture}\hfill\begin{tikzpicture}
	\begin{pgfonlayer}{nodelayer}
		\node [style=triangle] (19) at (-2.75, 0) {};
		\node [style=losange] (21) at (-2, -1.25) {};
		\node [style=triangle] (16) at (0, 0) {};
		\node [style=black] (0) at (0, 0) {};
		\node [style=none] (1) at (0, -0.5) {\tiny X};
		\node [style=none] (2) at (-1, 2) {};
		\node [style=none] (3) at (-1, -2) {};
		\node [style=none] (4) at (1, 2) {};
		\node [style=none] (5) at (1, -2) {};
		\node [style=black] (6) at (-2, 1) {};
		\node [style=black] (7) at (-2, -1.25) {};
		\node [style=black] (8) at (-2.75, 0) {};
		\node [style=black] (9) at (2, 1.25) {};
		\node [style=black] (10) at (2, -0.5) {};
		\node [style=losange] (11) at (3.25, 0) {};
		\node [style=black] (11) at (3.25, 0) {};
		\node [style=black] (12) at (3, -2) {};
		\node [style=none] (13) at (2, 0.75) {\tiny XY};
		\node [style=none] (14) at (2, -1) {\tiny XZ};
		\node [style=none] (15) at (3.25, -0.5) {\tiny XY};
		\node [style=none] (17) at (-2, -1.75) {\tiny Z};
		\node [style=none] (18) at (-2.75, -0.5) {\tiny Y};
		\node [style=none] (22) at (-2, 1.425) {\tiny XY};
		\node [style=none] (23) at (3, -1.55) {\tiny X};
	\end{pgfonlayer}
	\begin{pgfonlayer}{edgelayer}
		\draw [style=pointille] (2.center) to (3.center);
		\draw [style=pointille] (4.center) to (5.center);
		\draw (0) to (7);
		\draw (0) to (11);
		\draw (6) to (0);
		\draw (0) to (8);
		\draw (8) to (6);
	\end{pgfonlayer}
\end{tikzpicture}\hfill\begin{tikzpicture}
	\begin{pgfonlayer}{nodelayer}
		\node [style=glosange] (29) at (-3.5, -0.5) {};
		\node [style=grond] (28) at (0, 0) {};
		\node [style=losange] (27) at (2, 1.25) {};
		\node [style=grond] (24) at (-2.25, -1.5) {};
		\node [style=glosange] (26) at (-2.25, 0.75) {};
		\node [style=rond] (19) at (-2.25, 0.75) {};
		\node [style=losange] (21) at (-3.5, -0.5) {};
		\node [style=losange] (16) at (0, 0) {};
		\node [style=black] (0) at (0, 0) {};
		\node [style=none] (1) at (0, -0.6) {\tiny X};
		\node [style=none] (2) at (-1, 2) {};
		\node [style=none] (3) at (-1, -2) {};
		\node [style=none] (4) at (1, 2) {};
		\node [style=none] (5) at (1, -2) {};
		\node [style=black] (7) at (-3.5, -0.5) {};
		\node [style=black] (8) at (-2.25, 0.75) {};
		\node [style=black] (9) at (2, 1.25) {};
		\node [style=black] (10) at (-2.25, -1.5) {};
		\node [style=black] (12) at (2, -1) {};
		\node [style=none] (13) at (2, 0.75) {\tiny XY};
		\node [style=none] (14) at (-2.25, -2.1) {\tiny X};
		\node [style=none] (17) at (-3.5, -1.2) {\tiny Z};
		\node [style=none] (18) at (-2.25, 0.15) {\tiny XY};
		\node [style=none] (23) at (2, -1.55) {\tiny XY};
		\node [style=none] (30) at (-2.75, 1.25) {\footnotesize $v$};
	\end{pgfonlayer}
	\begin{pgfonlayer}{edgelayer}
		\draw [style=pointille] (2.center) to (3.center);
		\draw [style=pointille] (4.center) to (5.center);
		\draw (0) to (8);
		\draw (0) to (12);
		\draw (12) to (10);
		\draw (10) to (7);
		\draw (7) to (8);
		\draw (8) to (9);
	\end{pgfonlayer}
\end{tikzpicture}
\caption{We illustrate the 3 kinds of flow : GFlow on the left, Pauli Flow in the middle, and
Shadow Pauli Flow on the right. In each case, the correction required by the measurement of the central vertex 
involves the vertices with different Paulis represented with white shapes: triangle = $D\cap \odd
D$, diamond  = $\odd D \setminus D$ and circle= $D\setminus \odd D$. A dependency on the past (vertex
$v$) is allowed in the Shadow Pauli Flow case when it is compensated  by a  $D_v$  in the past
represented by grey shapes.}
\label{fig3}
\end{figure}

It leads to the following description of Pauli Flow: 
\begin{restatable}{proposition}{Pzero}\label{P0}

	Given $p:O^c \to 2^{I^c}$ and $\prec$ a strict partial order over $O^c$,

	$(p,\prec) \text{ is a Pauli Flow of } (G,I,O,\lambda)~\text{iff} $
	 $$ \forall u \in
	O^c,\,  u\in \cor({p(u)}) \text { and } 
	 \ana_{p,\prec}(u) =\emptyset$$ where  
	 $\ana_{p,\prec}(u):=\{v\in O^c,$ s.t.  $u\neq v$, $\neg (v\prec u)$  and  
	$u\in \infl({p(v))}\}$ is the set of vertices corrected after $u$ which correction has an ``anachronistic" impact on $u$.

\end{restatable}

In the rest of the paper, we will use the Proposition~\ref{P0} as a definition of Pauli Flow as it is simpler than the original one. The proof is given in the Appendix~\ref{proof:P0}.

This notion of \emph{impact} provides a set of constraints on the order according to which the qubits of the pattern can be measured: 
\begin{corollary}\label{cor:pf}

  $(G,I,O,\lambda)$ has a Pauli Flow iff there exists $p$ s.t. $K_p$ is a DAG\footnote{Directed Acyclic Graph}, where $K_p = (V(G), \left\{(u, v) \in (O^c)^2, v\in \infl(p(u)) 
  \right\})$.

  For such a $p$, $(p, \prec)$ will be a Pauli Flow for every partial order $\prec$ compatible with the DAG.

\end{corollary}

While the Pauli Flow condition is based on  the fixed-point properties of the underlying graph state
and the specific properties of Pauli measurements, we identify a third property, based on \emph{shadow correctors},  which allows
deterministic computation in MBQC\@. 

Indeed, an anachronous impact (i.e. an impact on an already measured qubit $v$) can actually be offset by a shadow corrector $D_v$ which is such that: (i) $D_v$ is correcting $v$; (ii) $D_v$ only acts on already measured qubits and $(iii)$ $D_v$ has only impacts on $v$.  Roughly speaking, a pattern $\mathcal P$ involving a plane measurement on $v$, say $YZ$, with angle $\alpha$ can be decomposed as a linear combination of two patterns $ \cos(\alpha)\mathcal P_Y+\sin(\alpha)\mathcal P_Z$ where $v$ is measured according to $Y$ (resp. $Z$)  in $\mathcal P_Y$ (resp. $\mathcal P_Z$).  If the anachronous impact on $v$ in $\mathcal P$ is say $Z$,  then it has actually no impact on $v$ in $\mathcal P_Z$, whereas,  in $\mathcal P_Y$, the action of the shadow corrector can turn the $Z$ action into a $Y$ action which has then no impact on $v$. 

Pauli Flow  can thus be extended to \emph{Shadow Pauli Flow}, in order to encounter shadow correctors:

\begin{definition}[Shadow Pauli Flow]\label{def:Extended Pauli Flow-strict}

	An open graph $(G,I,O,\lambda)$ has a Shadow Pauli Flow $(p,\prec)$, where $p:O^c \to 2^{I^c}$
	and $\prec$ partial order, if $\forall v\in O^c$, 
		\begin{itemize}
	\item {$v$ is corrected by $p(v)$, i.e.} $v\in \cor(p(v))$ and 
	\item 
	if $v$ is anachronistically impacted $(\ana_{p,\prec}(v)\neq \emptyset)$, then 
  $v$ is not a Pauli measurement $(|\lambda_v|=2)$, and it has a set of ``shadow correctors" 
	  $ {D_v}\subseteq I^c$ such that:
		\begin{itemize}
\item $v\in \cor (D_v)$ 
	and
\item $\forall u \in \ana_{p,\prec}(v),~\forall w\in ({D_v} \cup
	\odd{{D_v}})\setminus \{v\},~ w\notin \infl({D_v})$ 
	and $w\preccurlyeq u.$
		\end{itemize}
	\end{itemize} 
 \end{definition}

Note that for any vertex  $w \in O^c$, and for any set $D\subseteq V$, if  $w\in D\cup \odd{D}$ and $w\notin \infl({D})$ 
 then   $|\lambda_w|=1$.  This means that the new condition for Shadow Pauli
Flow implies that 
when the correction of $u$ impacts $v$
 that  has already been measured, $v$ must be
measured with a plane, and have an extra set of ``shadow correctors" $D_v$ such that  ${D_v} \cup
	\odd{{D_v}})\setminus \{v\}$  contains only vertices that are Pauli measured and that are in the
	past of any $u'$ which measurement has an anachronistic effect on $v$.

\begin{remark}\label{rk:input}
Similarly to GFlow and Pauli Flow, the existence of a Shadow Pauli Flow implies that any input is
measured in $XY$-plane or a subset of it.  
\end{remark}

Figure~\ref{fig:extPF} shows an open graph that has a Shadow Pauli Flow which is not a
Pauli Flow.

\begin{figure}

\begin{center}
\begin{tikzpicture}
	\begin{pgfonlayer}{nodelayer}
		\node [style=black] (8) at (-2, 0) {};
		\node [style=black] (10) at (-0.5, 0) {};
		\node [style=black] (11) at (1.75, 1.25) {};
		\node [style=black] (12) at (1, 0) {};
		\node [style=none] (13) at (-2, 0.5) {\tiny YZ};
		\node [style=none] (14) at (-0.5, 0.5) {\tiny XY};
		\node [style=none] (15) at (2.25, 1.5) {\tiny XY};
		\node [style=none] (16) at (-2, -0.5) {\tiny 0};
		\node [style=none] (17) at (1, 0.5) {\tiny X};
		\node [style=white] (18) at (1.75, -1.25) {};
		\node [style=none] (19) at (-0.5, -0.5) {\tiny 1};
		\node [style=none] (20) at (1, -0.5) {\tiny 2};
		\node [style=none] (21) at (2.25, 1) {\tiny 3};
		\node [style=none] (22) at (2.25, -1) {\tiny 4};
	\end{pgfonlayer}
	\begin{pgfonlayer}{edgelayer}
		\draw (8) to (10);
		\draw (10) to (12);
		\draw (10) to (11);
		\draw (11) to (18);
		\draw (10) to (18);
	\end{pgfonlayer}
\end{tikzpicture}
\caption{Example of open graph with Shadow Pauli Flow but no Pauli Flow with order $0\prec1\prec2\prec3$ and 
$p(0)=\{0,4\}, p(1)=\{4\}, p(2)=\{1\}, p(3)=\{4\}, \ana_{p,\prec}(1)=\{2,3\}, D_1=\{2\},  \ana_{p,\prec}(0)=\{2\}, D_0=\{0,2\}$}
\label{fig:extPF}
\end{center}

\end{figure}

\begin{theorem}[Main result]\label{thm:CNS}

	A pattern is robustly deterministic iff it is consistent with a Shadow Pauli Flow of its
	underlying open graph. 
	\end{theorem}

To prove Theorem~\ref{thm:CNS}, we first prove in the next Section (Proposition~\ref{thm:CSExtPauli}) that the existence of a Shadow
Pauli Flow is a sufficient condition for robust determinism. The necessity is proved in Proposition~\ref{thmCNExtPauli} section~\ref{sec:CN}.

We also show in Section~\ref{sec:algo} that Shadow Pauli Flow can be computed efficiently: 

{\begin{restatable}{theorem}{thmalgo}\label{thmalgo}
Given an open graph of order $n$, and a partial order over the non-output vertices of the graphs, a Shadow Pauli Flow compatible with the partial order, or a certificate of non existence, can be computed  in time polynomial in $n$. 
\end{restatable}}

\section{Shadow Pauli Flow is sufficient for Robust Determinism}\label{sec:CS}

\begin{proposition}\label{thm:CSExtPauli}

	Any pattern consistent with a Shadow Pauli Flow is robustly deterministic.

\end{proposition}

The rest of the section is dedicated to the proof of Proposition~\ref{thm:CSExtPauli}. 
The robustness of the consistent pattern relies on the fixed-point properties of underlying graph state and also the properties of the particular measurements.

Given a Shadow Pauli Flow $(p,\prec)$ of an open graph $(G,I,O,\lambda)$, the patterns $\cal P^{<,\alpha}$ consistent with $(p,\prec)$ are obtained as follows, by refining the partial order $\prec$ into a total order $<$ and by fixing the angles of measurements $\alpha_u$ for every $u\in O^c$. 

Indeed, given a vertex $u$, the total order $<$ induces 
 the following partition of the set vertices $v$ on which $p(u)$ is acting ($Act_v^{p(u)}\neq I$): 
\begin{itemize}
\item $A^u:=  \{v\in p(u)\cup \odd{p(u)}~|~v< u\wedge  |\lambda_v|=2\}$ the set of correctors which are already measured according to a plane;
\item $B^u:=  \{v\in p(u)\cup \odd{p(u)}~|~v< u\wedge |\lambda_v|=1\}$ the set of correctors which are already Pauli-measured; 
\item $C^u:=  \{v\in p(u)\cup \odd{p(u)}~|~u< v \text{ or } v \in O\}$, the set of correctors of $u$ which are not yet measured;
\item $F^u:= \{u\} \cap (p(u)\cup \odd{p(u)}$ completes the partition of $p(u)\cup \odd{p(u)}$. Notice that $F^u = \{u\}$ since the Shadow Pauli Flow conditions guarantees that $u\in
		p(u)\cup \odd{p(u)}$.
\end{itemize}

Moreover, for any set $S\in \{A^u, B^u, C^u, F^u\}$, we define $S_X := S \cap p(u)$ and  $S_Z:=
S\cap \odd{p(u)}$. Notice that $S_X \cup S_Z = S$.

The consistent pattern is then ${\cal P}^{<,\alpha}=\left(\prod_{u\in O^c}^< \X_{C^u_X}^{\s_u}\Z_{C^u_Z}^{\s_u}\M_u^{\lambda_u, \alpha_u}\right)\E_G\N_{I^c}$. 

The fundamental property of graph states gives us the first useful fixed-point property.

\begin{fact}

	For any $D\subseteq I^c$, and any $\ket \phi \in \mathbb C^{2^I}$, 
 	\begin{equation}\label{eq:fond}
		X_DZ_{\odd D} E_G \ket +_{I^c} \ket{\phi}_I = \pm  E_G \ket +_{I^c} \ket{\phi}_I
  	\end{equation}

\end{fact}

We also have the following standard fixed point property.

\begin{fact}

	For any $v\in O^c$, there exist $P_v, Q_v \in \lambda_v$, such that
	\begin{equation}\label{eq:FP}
	 (\cos(\alpha_v) P_v + \sin(\alpha_v) Q_v) \ket{+^{\lambda_v}_{\alpha_v}}=  \ket{+^{\lambda_v}_{\alpha_v}}
	\end{equation}

\end{fact}

Notice that if $|\lambda_v|=1$ then $P_v=Q_v$ and $\alpha_v=0\bmod \pi$,
thus $(\cos(\alpha_v) P_v + \sin(\alpha_v) Q_v)=\pm P_v$.

Then, we state a few  more almost fixed-point equations  which are then used to prove that the existence
of a Shadow Pauli Flow is a sufficient condition for robust determinism.

\begin{restatable}{lemma}{fixedPointBu}\label{fixedPoint:Bu}

	Given an open graph with a Shadow Pauli Flow, for any $u\in O^c$,
	\begin{equation}\label{eq:pauli}
		\left(\prod_{v\in B^u}  \bra{+^{\lambda_v}_{\alpha_v}}\right) X_{B^u_X} Z_{B^u_Z}  =
		\pm i^{| B^u_X\cap B^u_Z|}  \prod_{v\in B^u}  \bra{+^{\lambda_v}_{\alpha_v}}
	\end{equation}

\end{restatable}

The proof is given in the Appendix~\ref{proof:fixedPoint:Bu}.

\begin{restatable}{lemma}{fixedPointAu}\label{fixedPoint:Au}
	Given an open graph $(G,I,O,\lambda)$ with a Shadow Pauli Flow $(p,\prec)$, and  a  vertex $v$ with  non empty $\ana_{p,\prec}(v)$, 
	 then for any one-qubit Pauli operator $L_v$, and any multi-qubit Pauli operator $R$, $\exists \theta$ s.t. 
 	\begin{equation}\label{eq:useless3}
 	\left(\prod_{w\in {D_v}\cup \odd {D_v}}  \bra{+^{\lambda_w}_{\alpha_w}} \right) R E_G \ket
	+_{I^c} \ket{\phi}_I =
 	e^{i\theta}\left(\prod_{w\in {D_v}\cup \odd {D_v}}  \bra{+^{\lambda_w}_{\alpha_w}} \right)
	L_v R E_G \ket +_{I^c} \ket{\phi}_I
	\end{equation}
	where $D_v$ is as in the definition~\ref{def:Extended Pauli Flow-strict} of the Shadow Pauli Flow.

\end{restatable}

The proof is given in the Appendix~\ref{proof:fixedPoint:Au}.

~~\\
\noindent{\bf Proof of Proposition~\ref{thm:CSExtPauli}:}\\
We are now ready to prove by induction that the Shadow Pauli Flow is a sufficient condition for
robust determinism of $\cal P^{<,\alpha}$. By induction hypothesis, there exists
$r\in \mathbb C$ s.t.~the state before the measurement of $u$ is $\ket \psi = r \left(\prod_{v< u}
\bra{+^{\lambda_v}_{\alpha_v}}\right) E_G \ket +_{I^c} \ket{\phi}_I$.

The measurement of $u$ leads, up to a renormalisation, to either $\bra{+^{\lambda_u}_{\alpha_u}}\ket
\psi $ or $\bra{-^{\lambda_u}_{\alpha_u}}\ket \psi $. In the second case the correction
$X_{C^u_X}Z_{C_Z^u}$ is applied. Thus it is enough to show $\ket{\psi_0} =
\bra{+^{\lambda_u}_{\alpha_u}} \ket \psi $ and $ \ket{\psi_1} =
X_{C^u_X}Z_{C_Z^u}\bra{-^{\lambda_u}_{\alpha_u}}\ket \psi $ are equal up to a global phase, to show
strong determinism.
 
The Shadow Pauli Flow condition guarantees 
$u\in \cor({p(u)})$, which implies that for any $\alpha_u$, \begin{equation}\label{eq:corr}\bra{-^{\lambda_u}_{\alpha_u}} \simeq \bra{+^{\lambda_u}_{\alpha_u}} X_{F^u_X}  Z_{F_Z^u}\end{equation}
 
 As a consequence:
\[\begin{array}{rclr}
\ket{\psi_1} &\simeq& X_{C^u_X}Z_{C_Z^u}\bra{+^{\lambda_u}_{\alpha_u}}Z_{F^u_Z}X_{F^u_X} \ket \psi  &(Eq.~\ref{eq:corr})\\
	&\simeq&r X_{C^u_X}Z_{C_Z^u} (\prod\limits_{v\le u} \bra{+^{\lambda_v}_{\alpha_v}})
	Z_{F^u_Z}X_{F^u_X} E_G \ket +_{I^c} \ket{\phi}_I&\\
	&\simeq&r X_{C^u_X}Z_{C_Z^u} (\prod\limits_{v\le u} \bra{+^{\lambda_v}_{\alpha_v}}) X_{B^u_X}
	Z_{B^u_Z}Z_{F^u_Z}X_{F^u_X} E_G \ket +_{I^c} \ket{\phi}_I&(Eq.~\ref{eq:pauli})\\
	&\simeq&r X_{C^u_X}Z_{C_Z^u} (\prod\limits_{v\le u} \bra{+^{\lambda_v}_{\alpha_v}})
	X_{A^u_X}Z_{A^u_Z}X_{B^u_X} Z_{B^u_Z}Z_{F^u_Z}X_{F^u_X} E_G \ket +_{I^c} \ket{\phi}_I&~(Eq.~\ref{eq:useless3})\\ 
    &\simeq&r (\prod\limits_{v\le u} \bra{+^{\lambda_v}_{\alpha_v}}) X_{p(u)}Z_{\odd{p(u)}} E_G \ket
		+_{I^c} \ket{\phi}_I&\\  
		&\simeq&r (\prod\limits_{v\le u} \bra{+^{\lambda_v}_{\alpha_v}}) E_G \ket +_{I^c} \ket{\phi}_I\simeq \ket{\psi_0} &~(Eq.~\ref{eq:fond})
\end{array}\]
\hfill$\Box$

\section{Shadow Pauli Flow is necessary for robust determinism}\label{sec:CN}

The main result of this Section is the proof of the necessity condition of Shadow Pauli Flow.

\begin{restatable}{proposition}{thmpush}\label{thmCNExtPauli}
If a pattern $\mathcal P$ is robustly deterministic then it is consistent with a Shadow Pauli Flow of
its underlying open graph.
\end{restatable}

To prove this, we start by proving a useful property (Lemma~\ref{lem:meas}) of robustly
deterministic computations, that allows us to have necessity conditions for the GFlow when all
measurements are performed according to a plane (Proposition~\ref{gflowfix}, which fixes the
original proof from~\cite{browne2007generalized}),
then we prove that for patterns that have the Pauli measurements first, the Pauli Flow is necessary (Proposition~\ref{thm:PFrd}).
Finally, we prove that for general robustly deterministic patterns, Shadow Pauli Flow is a necessary condition.

\subsection{Necessity of GFlow for Pauli measurement-free patterns}\label{sec:NecGFlow}

The GFlow is a particular case of Pauli Flow, when all measurements are performed according to a plane (i.e. $\forall u\in O^c$, $|\lambda_u| =2$). The original proof from~\cite{perdrix2017determinism} showing the necessity of  GFlow used a wrong Lemma (Lemma 4 of~\cite{browne2007generalized}) that stated that if two
states were the same after any  measurement  of one qubit in a plane then they were the same.
 Indeed, as can be seen in Lemma~\ref{lem:meas} below, there is a case where the measured
qubit is separated from the rest and  is in a state that is an eigenvector of the  observable not
in the plane ($Z$ in the $XY$-plane measurement case). We show in Proposition~\ref{gflowfix} that the necessity of GFlow still holds in this particular
case as by robust determinism it can be corrected: indeed, as the $\bra{-}_\alpha$ branch can be changed to
the $\bra {+}_\alpha $ branch by applying precisely the observable not in the plane, it
cannot be an eigenvector. 

\begin{lemma}\label{lem:meas}

	Let $\lambda \subset \{X,Y,Z\}$ a set of two elements and $P$ be the Pauli not in $\lambda$.
	Given $\ket \phi $ and $\ket{\phi'}$ two states of a register $V$ of qubits, and  $u\in V$,
	if $\forall \alpha,~\bra{+^\lambda_\alpha}_u\ket {\phi} \simeq \bra {+^\lambda_\alpha}_u \ket
	{\phi'}$ and $|\bra{+^\lambda_\alpha}_u\ket {\phi}|=\frac 1{\sqrt 2}$ then $$\begin{cases}\ket \phi \simeq\ket {\phi'} \text{, or}&
	\\ \exists x\in \{0,1\}, \exists \ket{\psi}~\text{s.t.}~\ket \phi \simeq \ket
	{+^P_{x \pi}}_u \otimes \ket \psi~\text{and}~\ket{\phi'} \simeq  \ket {-^P_{x \pi}}_u \otimes \ket \psi&\end{cases}$$
\end{lemma}

\begin{proof} Since $\left\{\ket{+^P_0}_u,\ket{-^P_0}_u\right\}$ is a basis, $\ket{\phi}$ and $\ket{\phi'}$ can be written as $\ket{\phi}=\ket{+^P_0}_u\ket {\phi_+} +  \ket{-^P_0}_u\ket {\phi_-} $ and $\ket{\phi'}
=\ket{+^P_0}_u\ket {\phi'_+} +  \ket{-^P_0}_u\ket {\phi'_-} $ where $\ket {\phi_+}$, $\ket
{\phi_-}$, $\ket {\phi'_+}$ and $\ket {\phi'_-}$ are not normalized. Considering  $\alpha =0$, we have
$\ket{\phi_+}\simeq \ket {\phi'_+}$ and similarly when $\alpha = \pi$, $\ket{\phi_-}\simeq \ket
{\phi'_-}$. So $\exists \theta, \theta'$ s.t.
$\ket {\phi'} = e^{i\theta}\ket{+^P_0}_u\ket {\phi_+} +  e^{i\theta'}\ket{-^P_0}_u\ket {\phi_-}$.
When $\alpha =\pi/2$, it implies $\ket {\phi_+} + i\ket {\phi_-}\simeq e^{i\theta}\ket {\phi_+} +  e^{i\theta'}i\ket {\phi_-}$.
\begin{itemize}
\item  if $\ket {\phi_+}$ and $\ket {\phi_-}$ are not colinear, we have $\theta = \theta'$ and thus $\ket\phi \simeq \ket {\phi'}$.
\item Otherwise, since $|\bra{+^{\lambda}_0}_u\ket {\phi}|=|\bra{+^{\lambda}_\pi}_u\ket
	{\phi}|=\frac 1{\sqrt 2}$,  there exists $\gamma$ s.t. $\ket {\phi_+}=e^{i\gamma}\ket {\phi_-}$.
	Moreover, the condition $|\bra{+^{\lambda}_{\frac{\pi}{2}}}_u\ket {\phi}|=\frac 1{\sqrt 2}$
	implies $\gamma = 0\bmod \pi$. As a consequence there exist $ x\in \{0,1\}$, and $\ket{\psi}$
	s.t.  $\ket \phi = \ket{+_{x\pi}^{P}} \otimes \ket \psi$. Similarly,  there exist $ y\in \{0,1\}$, and
	$\ket{{\psi'}}$ s.t.  $\ket {\phi'} = \ket{+_{y\pi}^{P}} \otimes \ket {{\psi'}}$. If $x=y$ then $\ket\phi
		\simeq \ket {\phi'}$, otherwise $\ket {\phi'} \simeq \ket {-_{x\pi}^{P}} \otimes \ket \psi$.
\end{itemize}

\end{proof}

\begin{corollary}\label{lem:oneQbitEquiv}

	Let $\ket \phi $ a state of a register $V$ of qubits, $\lambda \subset \{X,Y,Z\}$ a set of two elements, $u\in V$, $C$ a multi-qubit Pauli  not acting on $u$ and $P\in \{\X,\Y, \Z\}\setminus  \lambda$. 
	If $\forall \alpha,~\bra{+^{\lambda}_\alpha}_u \ket {\phi} \simeq C \bra{-^{\lambda}_\alpha}_u \ket{\phi}$ 
	then $\ket \phi \simeq (P_u\otimes C) \ket \phi$. 

\end{corollary}

\begin{proof} 
Let $\ket{\phi'} = (P_u\otimes C) \ket{\phi}$. According to Lemma~\ref{lem:meas}, $\ket{\phi'}
\simeq \ket{\phi}$ or $\exists x\in \{0,1\}, \exists \ket{\psi}$ s.t.   $\ket \phi \simeq
\ket{+_{x\pi}^P}_u \otimes \ket \psi$ and $\ket{\phi'} \simeq \ket {-_{x\pi}^P}_u\otimes \ket \psi$. In
this second case, we would have $\ket{\phi'} = (P_u\otimes C) \ket{\phi}  \simeq (P_u\otimes C)
\ket{+_{x\pi}^P}_u \otimes \ket \psi \simeq \ket{+_{x\pi}^P}_u \otimes C\ket \psi$ since
$P\ket{+_{x\pi}^P} \simeq \ket{+_{x\pi}^P}$ which is a contradiction. 

\end{proof}

\begin{proposition}\label{gflowfix}

If $\mathcal P$ is robustly deterministic and has no Pauli measurement then the underlying open graph has a GFlow.

\end{proposition}

\begin{proof} Let $<$ be the order according to which the qubits are measured in the pattern $\mathcal P$.
For any $u\in O^c$, the robust determinism condition implies  the two states obtained after the measurement of $u$, for any angle $\alpha$, and the corresponding corrections are equal up to a global phase:
\[\forall \vec \beta \in [0,2\pi)^{\{v \in O^c~|~v< u\}},
	\bra{+^{\lambda_u}_\alpha}\bra{+^{\vec{\lambda}}_{\vec
\beta}} E_GN_{V\setminus I}\ket {\phi}_I \simeq
X_{C_X^u}Z_{C_Z^u}\bra{-^{\lambda_u}_\alpha}\bra{+^{\vec{\lambda}}_{\vec \beta}} E_GN_{V\setminus I}\ket {\phi}_I \]
where $\bra{+_{\vec{\beta}}^{\vec{\lambda}}}$ is the tensor product of
$\bra{+^{\lambda_v}_{\beta_v}}$ over the $v\in O^c$ s.t. $v< u$.

According to corollary~\ref{lem:oneQbitEquiv}, we have
\[\forall \vec \beta \in [0,2\pi)^{\{v \in O^c~|~v< u\}}, \bra{+^{\vec{\lambda}}_{\vec \beta}}
E_GN_{V\setminus I}\ket {\phi}_I \simeq  \bra{+^{\vec{\lambda}}_{\vec \beta}} (X_{C_X^u}Z_{C_Z^u}\otimes P_u) E_GN_{V\setminus I}\ket {\phi}_I\]
where $P\in \{\X,\Y,\Z\}\setminus \lambda_u$.

Let $\ket{\psi} = E_GN_{V\setminus I}\ket {\phi}_I$ and $\ket{\psi'}= (X_{C_X^u}Z_{C_Z^u}\otimes P_u) E_GN_{V\setminus I}\ket {\phi}_I$. By iterative applications of Lemma~\ref{lem:meas}, there exists $F\subseteq \{v \in O^c~|~v< u\}$ and $\forall w\in F, x_w\in \{0,1\}$, s.t. $\ket{\psi} = \left(\bigotimes_{w\in F}\ket{+_{x_w\pi}^{P_w}}_u\right)\otimes \ket {\Psi}$ and $\ket{\psi'} = \left(\bigotimes_{w\in F}\ket{-_{x_w\pi}^{P_w}}_w\right)\otimes \ket {\Psi}$, with $P_w\notin \lambda_w$.

Notice that $\ket{\psi'} \simeq (X_{C_X^u}Z_{C_Z^u}\otimes P_u)\ket \psi$, so $\left(\bigotimes_{w\in F}\ket{-_{x_w\pi}^{P_w}}_u\right)\otimes \ket {\Psi} \simeq \left(\bigotimes_{w\in F}\ket{+_{x_w\pi}^{P_w}}_w\right)\otimes (X_{C_X^u}Z_{C_Z^u}\otimes P_u) \ket {\Psi}$ since $u\notin F$ and $(C^u_X\cup C^u_Z)\cap F = \emptyset$. As a consequence $F=\emptyset$ and 

 \begin{equation}\label{eq:stab}E_GN_{V\setminus I}\ket {\phi}_I\simeq X_{C_X^u}Z_{C_Z^u}P_uE_GN_{V\setminus I}\ket {\phi}_I\end{equation}

  We show that $C^u_X \cap I=\emptyset$. Indeed, by contradiction, assume there exists $e\in I\cap C^u_X$. We initialize this qubit $e$  in the $\ket 0$-state so that  $E_GN_{V\setminus I}\ket {\phi}_I =\ket {0}_e\otimes \ket{\phi}$ for some $\ket{\phi}$, if $e\neq u$, then $X_{C_X^u}Z_{C_Z^u}P_u\ket {0}_e\otimes \ket{\phi} \simeq  \ket{1}_e\otimes X_{C_X^u\setminus \{e\}}Z_{C_Z^u\setminus \{e\}}P_u \ket{\phi}$ which contradicts Equation~\ref{eq:stab}. If $u=e$ then $\lambda_u = \{X,Y\}$, so $P_u=Z_u$ (see remark~\ref{rk:input}). As a consequence, we have $X_{C_X^u}Z_{C_Z^u}Z_u\ket {0}_e\otimes \ket{\phi} \simeq  \ket{1}_e\otimes X_{C_X^u\setminus \{e\}}Z_{C_Z^u\setminus \{e\}} \ket{\phi}$,  which also contradicts Equation~\ref{eq:stab}.

 Moreover, if $\ket {\phi}_I = \ket+_I$, then $E_GN_{V\setminus I}\ket {\phi}_I$ is nothing but the graph state $\ket G$. Hence $X_{C_X^u}Z_{C_Z^u}P_u$ is in the stabilizer of $\ket G$ so it must be of the form $X_D Z_{\odd{D}}$ for some $D$. We define $g(u)$ to be this set $D$, i.e. $g(u):= \begin{cases} C_X^u&\text{if  $\lambda_u=\{X,Y\}$}\\C_X^u\cup\{u\}&\text{otherwise}\end{cases}$. Notice that  $\odd{g(u)} = \begin{cases} C_Z^u&\text{if  $\lambda_u=\{Y,Z\}$}\\C_Z^u\cup\{u\}&\text{otherwise}\end{cases}$.

 We can show that $(g, <)$ is a GFlow, 
  Indeed all vertices of $g(u)\setminus \{u\}$ and $\odd{g(u)}\setminus \{u\}$ are larger than $u$.  Moreover if $\lambda_u=\{X,Y\}$ then $ u\notin g(u)$ and $u\in \odd{g(u)}$; if $\lambda_u=\{Y,Z\}$ then $ u\in g(u)$ and $u\notin \odd{g(u)}$; and if $\lambda_u=\{X,Z\}$ then $ u\in g(u)$ and $u\in \odd{g(u)}$. 
\end{proof}

\subsection{Necessity of Pauli Flow for Pauli first  measurement patterns}\label{sec:NecPauliFirst}

We consider here patterns where the Pauli measurements are done first and we prove that for this
family of patterns, the ``standard" Pauli Flow is a necessary condition for robust determinism. 
As $u$ is Pauli measured if and  only if $|\lambda_u|=1$, we formally define Pauli-first measurement patterns by:
\begin{definition}
A pattern $\left(\prod_{u\in O^c}^< \X_{\mathbf x(u)}^{\s_u}\Z_{\mathbf z(u)}^{\s_u}\M_u^{\lambda_u, \alpha_u}\right)\E_G\N_{I^c}$ is \emph{Pauli-first} if $|\lambda_u|<|\lambda_v| \Rightarrow u<v$.  
\end{definition}

In the following we give a characterisation of the stabilizers of any graph state partially measured
using Pauli observables.

\begin{definition}

	Given a set $A$, and a collection  $P_A=\{ r_i P_i\}_{ i\in A}$ where $P_i\in \{X,Y,Z\}$ and
	$r_i\in \{-1,1\}$ of Pauli observables, let $\bra {P_A} := \prod_i \bra{+^{P_i}_{r_i \pi}}$,
	$x(P_A):=\{i\in A~|~P_i= X\text{ or } Y\}$ and $z(P_A):=\{i\in A~|~P_i= Y\text{ or } Z\}$.

\end{definition}

\begin{lemma}\label{lem:meastab}

	Given a graph state $\ket G$, a Pauli operator $M$, and a collection  $P_A$ of Pauli observables
	s.t. $|\bra {P_A}\ket G| = 2^{\frac{-|A|}2}$, $M\bra {P_A}\ket G \simeq \bra {P_A}\ket G$ iff
$$\exists S\subseteq V(G), S\cap z(P_A) = \odd{S}\cap x(P_A)~\text{and}~M \simeq X_{S\setminus
x(P_A)} Z_{\odd{S}\setminus z(P_A)}.$$

\end{lemma}

\begin{remark}
Notice that the condition $S\cap z(P_A) = \odd{S}\cap x(P_A)$ is equivalent to say that for all $u$ already measured (i.e. $u\in A$),
$$\begin{cases}
	\text{ if } u\in S\setminus \odd S \text{ then } u  \text{ has been X-measured i.e. } u\in x(P_A)\setminus z(P_A) ,
	\\\text{  if } u\in S\cap \odd S \text{ then } u \text{ has been Y-measured i.e. } u\in x(P_A)\cap z(P_A), \text{  and}
	\\ \text{ if  } u\in (\odd S\setminus S) \text{ then } u \text{ has been Z-measured i.e. } u\in z(P_A)\setminus x(P_A). 
 \end{cases}$$
\end{remark}

\begin{proof}
($\Leftarrow$) First notice that  $ S\cap z(P_A) = \odd{S}\cap x(P_A) $ implies  $S\cap z(P_A) \subseteq x(P_A)$, so $S\setminus x(P_A)\subseteq V\setminus (z(P_A)\cup x(P_A)) = V\setminus A$. Similarly,  $\odd{S}\setminus z(P_A)\subseteq V\setminus A$. 
\begin{eqnarray*}
X_{S\setminus x(P_A)} Z_{\odd{S}\setminus z(P_A)}\bra {P_A}\ket G &=&\bra {P_A}X_{S\setminus x(P_A)} Z_{\odd{S}\setminus z(P_A)}\ket G \\
&\simeq &\bra {P_A}X_{S\cap (x(P_A)\setminus z(P_A))} Z_{\odd{S}\cap(z(P_A)\setminus x(P_A)) } \\
&&  (ZX)_{ S\cap \odd{S} \cap x(P_A) \cap z(P_A)} X_{S\setminus x(P_A)} Z_{\odd S\setminus z(P_A)}\ket G \\
&\simeq&\bra {P_A}X_{S} Z_{\odd{S}}\ket G 
\end{eqnarray*}
\noindent ($\Rightarrow$) By induction on the size of $A$. If $A=\emptyset$ the property is true. Assume the property is true for $A$ of size $k$, and let $u\in V\setminus A$. W.l.o.g.~assume $u$ is X-measured, and $|\bra {+_u}\bra{P_A}\ket G| = 2^{\frac{-(|A|+1)}2}$.
As $|\bra {+_u}\bra{P_A}\ket G| =  \frac{|\bra{P_A}\ket G| }{\sqrt 2}$,  $\bra{P_A}\ket G$ is not an
eigenvector of $X_u$, so there exists a Pauli operator $M'$ s.t. $M'\bra{P_A}\ket G \simeq
\bra{P_A}\ket G$ and $M'$ anticommutes with $X_u$. By induction hypothesis, there exists $S_0$ s.t. $S_0\cap z(P_A)=\odd{S_0}\cap x(P_A)$ and $M' = X_{S_0\setminus x(P_A)} Z_{\odd{S_0}\setminus z(P_A)}$. Moreover, since $M'$ and $X_u$ anticommute, we have $u\in \odd{S_0}$. 

Notice that $\bra {P_A}\ket G = \ket +_u \otimes \bra +_u\bra {P_A}\ket G + \ket -_u \otimes \bra -_u\bra {P_A}\ket G$. Moreover,  $\bra -_u\bra {P_A}\ket G \simeq \bra -_uX_{S_0\setminus x(P_A)} Z_{\odd{S_0}\setminus z(P_A)}\bra {P_A}\ket G \simeq  X_{S_0\setminus x(P_A)\setminus u} Z_{\odd{S_0}\setminus
z(P_A)\setminus u}\bra +_u\bra {P_A}\ket G $, since $u\in \odd{S_0}$. 

As a consequence, $$\bra {P_A}\ket G = \ket +_u \otimes \bra +_u\bra {P_A}\ket G + e^{i\theta}\ket -_u\otimes \X_{(S_0\setminus x(P_A))\setminus u} Z_{(\odd{S_0}\setminus
z(P_A))\setminus u}\bra +_u\bra {P_A}\ket G $$

Let $M$ be a Pauli operator s.t. $M \bra +_u\bra {P_A}\ket G \simeq \bra +_u\bra {P_A}\ket G $.
Notice that $\bra {P_A}\ket G $ is a eigenvector of $X_u^m\otimes M$ where
$$m=\begin{cases}0&\text{if $M$
and $X_{(S_0\setminus x(P_A))\setminus u} Z_{(\odd{S_0}\setminus z(P_A))\setminus u}$ commute}\\
1&\text{if they anticommute}\end{cases}.$$ 

Thus by induction hypothesis there exists $S_1$ s.t.
$X^m_u\otimes M \simeq X_{S_1\setminus x(P_A)} Z_{\odd{S_1}\setminus z(P_A)}$.

Notice that $m=1 \iff u\in {S_1}$ so $M= X_{S_1\setminus x(P_A)} Z_{\odd{S_1}\setminus z(P_A)}X^m_u
= X_{S_1\setminus x(P_{A\cup \{u\}})} Z_{\odd{S_1}\setminus z(P_{A\cup \{u\}})}$.

\end{proof}

This allows to prove the following proposition  (see appendix~\ref{anxthm:PFrd}): 
\begin{restatable}{proposition}{thmPFrd}\label{thm:PFrd}
A robust deterministic Pauli-first pattern has a Pauli Flow.
\end{restatable}

{A direct corollary of Proposition~\ref{thm:PFrd} is that an open graph can be used to performed a deterministic computation if and only if it admits a Pauli Flow: 
 \begin{theorem}\label{thm:PFCNS}
An open graph can be used to perform a robustly deterministic MBQC if and only if it has a Pauli Flow. 
 \end{theorem}}

\subsection{Necessity of Shadow Pauli Flow}\label{sec:NecExtended Pauli Flow}

\subsubsection{Pushing the Pauli measurements first in robust deterministic patterns}

In early papers~\cite{raussendorf2000quantum}, it is mentioned that Pauli measurements can be done at the beginning of the computation, however it has never been proved that this transformation preserves robust determinism.
 We discuss in this section how Pauli measurement can be pushed to the beginning 
  in robustly deterministic MBQC patterns preserving the robust determinism. 

 First we define the relation  $\rhd_u$,   such that ${\mathcal P}_0 \rhd_u {\mathcal P}_1$ if ${\mathcal P}_1$ can be obtained from  ${\mathcal P}_0$ by bringing one step forward the Pauli measurement of $u$ in 
 ${\mathcal P}_0$.  
 
\begin{definition}\label{defpush} The relation $\rhd_u $ over patterns  is inductively defined as 
\begin{itemize}
\item $\X^{\s_v}_A\Z^{\s_v}_B\M_v^{\lambda, \alpha}\mathcal P_0
	\rhd_u\X^{\s_v}_A\Z^{\s_v}_B\M_v^{\lambda, \alpha}  \mathcal P_1$ if $\mathcal P_0 \rhd_u \mathcal
	P_1$ $v\neq u$
\item $C^{s_u}\M_u^{P,\alpha}\sigma^{s_v}_uR^{s_v}\M_v^{\{Q,Q'\}, \theta} \mathcal P  \rhd_u
	R'^{s_v}\M_v^{\{Q,Q'\}, \theta} C'^{s_u}\M_u^{P,\alpha} \mathcal P$
\[\text{with} \begin{cases}   
~\text{If}~\sigma P=- P \sigma &  R'=CR~\text{and}~C'=C \\
~\text{Otherwise if $RC=CR$ xor $\sigma=\widetilde P$} &R'=R~\text{and}~C'=C\\
~\text{Otherwise}~  &R'\in\{R,\emptyset\}~\text{and}~C'\in\{\widetilde {Q_v}C,\widetilde {Q'_v}C\}\\
 \end{cases}\] 
 where $R,C$ are (multi-qubit) Pauli correctors which do not act on $u$; $P, Q, Q'\in \{\X,\Y,\Z\}$;
 $Q\neq Q'$; $\sigma \in \{I,\X,\Z,\X\Z\}$\footnote{with a slight abuse of notation $\sigma=I$ means that the correctors of $v$ do not act on $u$.}
 and $\widetilde Q:= \begin{cases}\X\Z&\text{if $Q= \Y$}\\ Q &\text{otherwise}\end{cases}$. 
\end{itemize}
\end{definition}

 We say that $\mathcal P \rhd \mathcal P'$ if there exist $u_1,\ldots, u_k$ such that $\mathcal P \rhd_{u_1}\ldots \rhd_{u_k} \mathcal P'$.

\begin{lemma}
\label{lem:push}
For any pattern $\mathcal P$, if there exists no patterns $\mathcal P'$ such that $\mathcal P\rhd \mathcal P'$, then $ \mathcal P$ is  Pauli-first.
\end{lemma}

\begin{proof}
	Let $u$ be the last vertex with a Pauli measurement following a non Pauli measurement on a vertex $v$,
	on some pattern $\mathcal P$, following the second case of definition~\ref{defpush} one can always find
	$R'$ and $C'$ to define $\mathcal P'$ satisfying $\mathcal P \rhd_u \mathcal P'$.
\end{proof}

\begin{restatable}{proposition}{Paulifirstpattern}\label{Paulifirstpattern}
 If $\mathcal P_0 \rhd \mathcal P_1$ and $\mathcal P_0$ is robustly deterministic so is $\mathcal P_1$. Moreover, $\interp {\mathcal P_0} = \interp {\mathcal P_1} $.
\end{restatable}

The proof of this proposition is given in Appendix~\ref{appendixthm}.

Notice that the converse of Proposition~\ref{Paulifirstpattern}  is not true:  it is possible to turn a non robustly  deterministic pattern  into a robustly deterministic one by doing the Pauli measurements first. 
 Indeed, the patterns  $Z_3^{s_2}M_2^ZZ_2^{s_1}M_1^{Y,Z} E_{1,2}E_{2,3}N_1N_2N_3$ and $Z_3^{s_2}M_2^ZZ_3^{s_1}Z_2^{s_1}M_1^{Y,Z} E_{1,2}E_{2,3}N_1N_2N_3$ both lead to  $M_1^{Y,Z} Z_1^{s_2} Z_3^{s_2} M^Z_2E_{1,2}E_{2,3}N_1N_2N_3$  which is robustly deterministic when the Pauli measurement is performed first, however only the former is robustly deterministic.

\subsubsection{Robust deterministic measurement patterns have Shadow Pauli Flow}

The  proof that having a Shadow Pauli Flow is necessary, is given 
 in  Appendix~\ref{anxthm:push} and relies on the fact  
 that if all patterns obtained by pushing forward a Pauli measurement have a Shadow Pauli Flow, so does the original pattern and we can build a compatible Shadow Pauli Flow using the ones where the Pauli measurements are pushed forward.

 Indeed, combining Proposition~\ref{Paulifirstpattern} and Lemma~\ref{lem:push}, from any robustly deterministic pattern one can build a
 robustly deterministic pattern in which the Pauli measurements are done before the general measurements.

Then,  given a robustly deterministic pattern $\mathcal P$, Lemma~\ref{lem:push} guarantees that by
pushing the Pauli measures in  $\mathcal P$ one obtains  Pauli-first patterns $\mathcal P'$  with
$\mathcal P \rhd  \mathcal P'$. Proposition~\ref{Paulifirstpattern} ensures that each $\mathcal P'$  is
robustly deterministic, and therefore  is consistent with a Pauli Flow by  Proposition~\ref{thm:PFrd}.
By induction we show in Appendix~\ref{anxthm:push} that $\mathcal P$ is consistent with a Shadow Pauli Flow.

\section{Computing Shadow Pauli Flow}\label{sec:algo}

{Like GFlow and Pauli Flow, Shadow Pauli Flow can be computed efficiently: 
\thmalgo*

To prove Theorem~\ref{thmalgo}, we consider the following algorithm:
Given an open graph $(G,I,O, \lambda)$ and a partial order $\prec$ of $O^c$, first it is convenient
to extend $\prec$ to $V$ s.t.\ any vertex in $O$ is not smaller than any vertex in $O^c$. 
The algorithm consists in pre-processing all the shadow correctors: for each vertex $v$ such that
$|\lambda(v)|=2$, find, if it exists a \emph{shadow corrector} $D_v$ s.t. $v\in \cor(D_v)$ and for
any $u\in (D_v\cup Odd(D_v))\setminus \{v\}$, $\lambda(u) = Act_u^{D_v}$, so in particular
$|\lambda(u)|=1$. 
Such a $D_v$ is a subset of $\lambda^{-1}(\{X\})\cup \lambda^{-1}(\{Y\})\cup\{v\}$ s.t.
$Odd(D_v)\setminus \lambda^{-1}(\{Z\})\subseteq \{v\}$, the action on $v$ depending on its
measurement plane, e.g. $v\in D_v$ and $v\in Odd(D_v)$ when $\lambda(v)=\{X,Z\}$.
Thus, following the algebraic approach for flow \cite{mhalla2014graph,mitosek2024algebraic}, for a
fixed $v$, $D_v$ can be found in $O(n^3)$ time, if it exists, by solving the linear system $M.D= b$
with the additional linear constraint $D[v]=1$ (i.e. $v\in D$) when $Z\in \lambda(v)$ and $D[v]=0$
(i.e. $v\notin D$) otherwise, where the column vector $b$  and the matrix $M$ is obtained from the
adjacency matrix $\Gamma$ of $G$ by adding $1$ to the diagonal on $Y$-measured vertices and then
keeping only the columns that correspond to $v$ and to the vertices that are X or Y measured;  and
only the rows that are not $Z$ measured, thus $M =
(\Gamma+Id_{\lambda^{-1}(\{Y\})})[\lambda^{-1}(\{X\})\cup\lambda^{-1}(\{Y\})\cup\{v\} , V\setminus
\lambda^{-1}(\{Z\})]$.  $b$ is the zero vector when $\lambda (v) = YZ$ and has a single non-zero
entry $b[v]=1$ otherwise. One can double check that $D_v$ is a shadow corrector iff it is a solution
to this linear system. 

Thus in time $O(n.n^3)$ one can associate with all plane-measured vertex a shadow corrector when
it exists. Notice that if $(G,I,O, \lambda)$ has a Shadow Pauli Flow then each vertex has at most
one shadow corrector. Indeed if $D$ and $D'$ are two distinct shadow correctors of $v$, then
$X_{\tilde D} Z_{Odd(\tilde D)}$ is a stabilizer of the graph where $\tilde D = D\Delta D'$,
moreover for each $u\in \tilde D \cup Odd(\tilde D)$, $\lambda(u)=Act_u^D$, thus the last qubit of
this set to be measured will lead to a deterministic outcome which contradicts that each measurement
is balanced. 

In the second part of the algorithm, the corrector function $p:O^c\to I^c$ is constructed vertex by
vertex. For any $u$ in $O^c$, the set of correctors $p(u)$ is obtained as the solution of the linear
system $N.D=b$, where $N$ is the matrix
$N=(\Gamma+Id_{\lambda^{-1}(Y)})[\lambda^{-1}(X)\cup\lambda^{-1}(Y)\cup\{v\} \cup F_u\cup S_u,
V\setminus (\lambda^{-1}(Z)\cup F_u\cup S_u)]$, where $F_u$ is the \emph{future} of $u$, i.e.\ the
set $F_u:= \{w~|~u\prec w\}$ of vertices that are larger than $u$, and $S_u$ is the set of vertices
that have shadow correctors when $u$ is measured, i.e. $S_u=\{v~|~\forall w\in D_v, w\prec u \text{
or } w=u\}$. The column vector $b$ is the zero vector when $\lambda (v) = YZ$ and has a single
non-zero entry $b[v]=1$ otherwise. 
Solving such a linear system find correctors $p(u)$ that freely uses unmeasured qubits at this step
of the computation, together with shadow corrected ones, and possibly also Pauli measured qubits
when their action is compatible with their Pauli basis. 

The second part of the algorithm also has a $O(n^4)$ complexity. Notice that this algorithm can be seen as a generalisation of Pauli Flow algorithms, see for instance  \cite{mitosek2024algebraic}, with a pre-processing to identify shadow corrected vertices, and then a layer by layer algorithm that treat shadow corrected vertices as output as long as there shadow correctors are actually in the measured subset of vertices. 
}

\section{Conclusion}

In this paper, we  provide a necessary and sufficient condition for robust determinism by defining the Shadow
Pauli Flow, that can be efficiently computed. We have  also shown that any Shadow Pauli Flow can be changed to a Pauli Flow by
pushing forward the Pauli measurements.

Future work could be to consider the case where multiple qubits are measured simultaneously.
Indeed, robust determinism implies that qubits are measured one by one.

Finally another perspective is to investigate the applications of the Shadow Pauli Flow in the context of graphical languages like the ZX-calculus where GFlow and Pauli Flow are already extensively used.

\subsection*{Acknowledgments}

The authors would like to thank anonymous referees for their valuable comments on a previous version of this paper. This work is supported by the Plan France 2030 through the PEPR integrated
  project EPiQ (ANR-22-PETQ-0007) and the HQI platform (ANR-22-PNCQ-0002); by
  the European Union through the MSCA Staff Exchanges project QCOMICAL (Grant
  Agreement ID: 101182520); and by the Maison du Quantique MaQuEst.

\printbibliography

\newpage

\appendix

\section{Proof of Proposition~\ref{P0}}\label{proof:P0}

\Pzero*

	From the proof of property 1 in~\cite{perdrix2017determinism} we know that $(p, \prec)$ is a Pauli Flow for $(G, I, O,
	\lambda)$ if the following conditions hold
\begin{eqnarray*}
		(c_X)\quad X\in \lambda_v&\Rightarrow & v\in \odd{p(v)}\setminus \left(  \bigcup_{\substack{\neg(u \prec v) \\ u \notin O\cup\{v\}}} \odd{p(u)} \right)\\
(c_Y)\quad Y\in \lambda_v&\Rightarrow & v\in (\odd{p(v)}\Delta p(v))\setminus \left(  \bigcup_{\substack{\neg(u \prec v) \\ u \notin O\cup\{v\}}} \odd{p(u)}\Delta p(u) \right)\\
(c_Z)\quad Z\in \lambda_v&\Rightarrow & v\in {p(v)}\setminus \left(  \bigcup_{\substack{\neg(u \prec v) \\ u \notin O\cup\{v\}}} {p(u)} \right)
\end{eqnarray*}
Where $\Delta$ is the symmetric difference.

Notice that  $v\in \odd{p(v)}\setminus \left(  \bigcup_{\substack{\neg(u \prec v) \\ u \notin O\cup\{v\}}} \odd{p(u)} \right)$ means that $v\in \odd{p(v)}$ and $\forall u \in O^c\setminus \{v\}$ s.t. $\neg(u \prec v), v\notin \odd{p(u)}$.

Therefore, satisfying  $(c_X)$, $(c_Y)$ and $(c_Z)$ is equivalent to satisfy $(i)$ and $(ii)$, where
	\begin{itemize}
		\item 
		$(i)$ $\forall u \in O^c$
			\begin{itemize}
				\item  If  $X\in \lambda_u$ then $Act^{p(u)}_u\in \{Y,Z\}$ and 
				\item  If  $Y\in \lambda_u$ then $Act^{p(u)}_u\in \{X,Z\}$ and 
				\item  If  $Z\in \lambda_u$ then $Act^{p(u)}_u\in \{X,Y\}$
			\end{itemize}
			This corresponds to $u\in \cor (p(u))=\{v,  \forall P \in \lambda_u, [Act^{p(u)}_v ,P]\neq 0 \}$. 

		\item $(ii)$ $\forall (u, v) \in (O^c)^2;~\neg(u \prec v)~\mbox{and}~v \neq u$,
			\begin{itemize}
			\item  If  $X\in \lambda_v$ then $ Act^{p(u)}_v \notin \{Z,Y\}$    and 
			\item  If  $Y\in \lambda_v$ then $ Act^{p(u)}_v \notin \{X,Z\}$    and 
			\item  If  $Z\in \lambda_v$ then $ Act^{p(u)}_v \notin \{X,Y\}$    
	\end{itemize}
			This corresponds to  $\ana_{p,\prec} (u)=\emptyset$. 
\end{itemize}

\section{Proof of lemma~\ref{fixedPoint:Bu}}\label{proof:fixedPoint:Bu}

\fixedPointBu*
Recall that  $B^u:=  \{v\in p(u)\cup \odd{p(u)}~|~v< u\wedge |\lambda_v|=1\}$ is  the set of correctors which are already Pauli-measured, 
 $B^u_X := B \cap p(u)$ and  $B^u_Z:=
B\cap \odd{p(u)}$. 
	
	For any $v\in B^u$,
	 the Shadow Pauli Flow condition
implies $v\notin \infl(p(u))$. 
	So,
	\begin{itemize}
		\item if $\lambda_v =X$ then $v\notin \odd{p(u)}$, and then $v\in B_X^u \setminus B_Z^u$,
		\item if $\lambda_v = Y$ then $v\notin \odd{p(u)}\Delta p(u)$, and then $v\in B^u_X \cap B^u_Z$,
		\item if $\lambda_v = Z$ then $v\notin p(u)$, and then $v\in B_Z^u \setminus B_X^u$.
	\end{itemize}
	As a consequence, the Pauli operator
	$X_{B^u_X}Z_{B^u_Z} = (-i)^{| B^u_X\cap B^u_Z|} X_{B_X^u \setminus B_Z^u}Y_{B_Z^u \cap B_X^u}Z_{B_Z^u \setminus B_X^u}$
	can be absorbed by the projectors:  
	$$\left(\prod_{v\in B^u}  \bra{+^{\lambda_v}_{\alpha_v}}\right)      X_{B^u_X} Z_{B^u_Z}
	= \pm    i^{| B^u_X\cap B^u_Z|}  \prod_{v\in B^u}  \bra{+^{\lambda_v}_{\alpha_v}}$$

\section{Proof of lemma~\ref{fixedPoint:Au}}\label{proof:fixedPoint:Au}

\fixedPointAu*
	
	Given a vertex $v$ with  non-empty $\ana_{p,\prec}(v)$, the definition of the Shadow Pauli Flow
condition implies that  for any $w\in (D_v\cup \odd{D_v})\setminus \{v\}$, $|\lambda_w|=1$. Moreover, $w\notin \infl(D_v))$, therefore:
	\begin{itemize}
		\item $\lambda_w = X \Rightarrow w\in {D_v}\setminus \odd {D_v} $,
		\item $\lambda_w = Y \Rightarrow w\in {D_v} \cap \odd {D_v}$,
		\item $\lambda_w = Z \Rightarrow w\in \odd {D_v}\setminus {D_v}$.
	\end{itemize}
	As a consequence,
\begin{equation}\label{eq:pauli_UL}
\begin{aligned}
\left(\prod_{w\in ({D_v}\cup \odd {D_v})\setminus \{v\}} \bra{+^{\lambda_w}_{\alpha_w}} \right)
X_{{D_v}\setminus \{v\}} Z_{\odd {D_v}\setminus \{v\}} =\\
\pm i^{|({D_v}\setminus \{v\})\cap (\odd {D_v}\setminus \{v\}) |} \prod_{w\in ({D_v}\cup
\odd{D_v})\setminus \{v\}} \bra{+^{\lambda_w}_{\alpha_w}}
\end{aligned}
\end{equation}
	Like in Eq.~\ref{eq:pauli}, the complex number in Eq.~\ref{eq:pauli_UL} witnesses the fact
	that when $\lambda_w=Y$, we use $X_wZ_w$ rather than $Y_w = iX_wZ_w$.

	Now, the Shadow Pauli Flow condition guarantees also that
	 $v\in \cor(D_v)$, which 
	means that:
	\begin{itemize}
		\item $\lambda_v = \{X,Y\} \Rightarrow v\in \odd {D_v} \setminus {D_v}$,
		\item $\lambda_v = \{X,Z\} \Rightarrow v\in \odd {D_v} \cap {D_v}$,
		\item $\lambda_v = \{Y,Z\} \Rightarrow v\in  {D_v} \setminus \odd {D_v}$.
	\end{itemize}
	Therefore, the previous fixed-point property (Eq.~\ref{eq:pauli_UL}) can be completed to
	get:
	\begin{equation}\label{eq:useless}
	\left(\prod_{w\in ({D_v}\cup \odd {D_v})\setminus \{v\}} \bra{+^{\lambda_w}_{\alpha_w}} \right)
	X_{{D_v}} Z_{\odd {D_v}} =
	\pm \left(\prod_{w\in ({D_v}\cup \odd {D_v})\setminus \{v\}}
	\bra{+^{\lambda_w}_{\alpha_w}}\right) \lambda_v^\perp
	\end{equation}
	where $\lambda_v^\perp = \begin{cases}
		Z_v & \text{if $\lambda_v = \{X,Y\}$}\\
		Y_v = i X_vZ_v & \text{if $\lambda_v = \{X,Z\}$}\\
		X_v & \text{if $\lambda_v = \{Y,Z\}$}\end{cases}$.  

	Notice that the term $i^{|({D_v}\setminus \{v\})\cap (\odd {D_v}\setminus \{v\}) |}$ of Eq.~\ref{eq:pauli_UL} is not present in Eq.~\ref{eq:useless} has it is absorbed by
	$\lambda_v^\perp$. Indeed, for any $D$, $|D\cap \odd D|=0 \bmod 2$, thus $|({D_v}\setminus
	\{v\})\cap (\odd {D_v}\setminus \{v\}) | =1 \bmod 2$ iff $v\in {D_v}\cap \odd {D_v}$, i.e.
	$\lambda_v = \{X,Z\}$.

	Thus by combining Eq.~\ref{eq:useless} and~\ref{eq:fond}, we get that for any $v\in A^u$,
	and any (multi-qubit) Pauli operator $R$,
	\begin{equation}\label{eq:useless2}
	\left(\prod_{w\in {D_v}\cup \odd {D_v}}  \bra{+^{\lambda_w}_{\alpha_w}} \right)
	R E_G \ket +_{I^c} \ket{\phi}_I =
	\pm  \left(\prod_{w\in {D_v}\cup \odd {D_v}} \bra{+^{\lambda_w}_{\alpha_w}} \right)
	\lambda_v^\perp R E_G \ket +_{I^c} \ket{\phi_I}
	\end{equation}

	We additionally have,  according to Eq.~\ref{eq:FP}, for some $P_v,Q_v \in \lambda_v$. 
\begin{equation}
\begin{aligned}
	& \left(\prod\limits_{w\in {D_v}\cup \odd {D_v}}  \bra{+^{\lambda_w}_{\alpha_w}} \right) R E_G \ket
	+_{I^c} \ket{\phi}_I \\ 
	& = \left(\prod\limits_{w\in {D_v}\cup \odd {D_v}}  \bra{+^{\lambda_w}_{\alpha_w}} \right)
	(\cos(\alpha_v) P_v + \sin(\alpha_v) Q_v)R E_G \ket +_{I^c} \ket{\phi}_I \\
	& = \cos(\alpha_v)  \left(\prod\limits_{w\in {D_v}\cup \odd {D_v}}
	\bra{+^{\lambda_w}_{\alpha_w}} \right) P_v R E_G \ket +_{I^c} \ket{\phi}_I\\
	& + \sin(\alpha_v) \left(\prod\limits_{w\in {D_v}\cup \odd {D_v}}  \bra{+^{\lambda_w}_{\alpha_w}}
	\right) Q_v R E_G \ket +_{I^c} \ket{\phi}_I
\end{aligned}
\end{equation}

	Thanks to Equation~\ref{eq:useless}, we have
	$$\left(\prod\limits_{w\in {D_v}\cup \odd {D_v}} \bra{+^{\lambda_w}_{\alpha_w}} \right) Q_v R
	E_G \ket +_{I^c} \ket{\phi}_I = \pm\left(\prod\limits_{w\in {D_v}\cup \odd {D_v}}
	\bra{+^{\lambda_w}_{\alpha_w}} \right)\lambda_v^\perp Q_v R E_G \ket +_{I^c} \ket{\phi}_I$$

	Moreover, since $\lambda_v = \{P_v,Q_v\}$, we have $\lambda_v^\perp Q = \pm i P$. Thus,
	$$\left(\prod\limits_{w\in {D_v}\cup \odd {D_v}} \bra{+^{\lambda_w}_{\alpha_w}}
	\right)\lambda_v^\perp Q_v R E_G \ket +_{I^c} \ket{\phi}_I = \pm i \left(\prod\limits_{w\in
	{D_v}\cup \odd {D_v}} \bra{+^{\lambda_w}_{\alpha_w}} \right) P_v R E_G \ket +_{I^c}
	\ket{\phi}_I$$
As a consequence,
$$\begin{aligned}
&\left(\prod\limits_{w\in {D_v}\cup \odd {D_v}}  \bra{+^{\lambda_w}_{\alpha_w}} \right) R E_G \ket
+_{I^c} \ket{\phi}_I  =\\
&(\cos(\alpha_v)\pm i \sin(\alpha_v))  \left(\prod\limits_{w\in {D_v}\cup \odd {D_v}}
\bra{+^{\lambda_w}_{\alpha_w}} \right) P_v R E_G \ket +_{I^c} \ket{\phi}_I
\end{aligned}$$
	Hence, as  $\ana_{p,\prec}(v)\neq \emptyset$, it can absorb, up to a global phase, one of the
	Pauli operators of the plane $\lambda_v$. Combined with equation~\ref{eq:useless2}, any Pauli
	operator on such a vertex $v$ can be absorbed, and we get Eq~\ref{eq:useless3}: for any one-qubit
	Pauli operator $L_v$, and any multi-qubit Pauli operator $R$, $\exists \theta$ s.t.
	$$\left(\prod_{w\in {D_v}\cup \odd {D_v}} \bra{+^{\lambda_w}_{\alpha_w}} \right) R E_G \ket
	+_{I^c} \ket{\phi}_I =
	e^{i\theta}\left(\prod_{w\in {D_v}\cup \odd {D_v}}  \bra{+^{\lambda_w}_{\alpha_w}} \right)
	L_v R E_G \ket +_{I^c} \ket{\phi}_I$$

\section{Proof of Proposition~\ref{thm:PFrd}}\label{anxthm:PFrd}

\thmPFrd*

\begin{proof}
 $$\mathcal P' = \left(\prod_{u\in O^c~\text{s.t.}~|\lambda_u|=2}^\prec
 C_u^{s_u}M_u^{\lambda_u,\alpha_u}\right)\left(\prod_{u\in O^c~\text{s.t.}~|\lambda_u|=1}^\prec C_u^{s_u}M_u^{\lambda_u,\beta_u}\right)E_GN_{V\setminus I}$$
 
 Let $\bra{\vec \alpha} :=\bigotimes_{u\in O^c~\text{s.t.}~|\lambda_u|=2} \bra{+^{\lambda_u}_{\alpha_u}}$. Notice that if $\forall u\in O^c, |\lambda_u|=1$, then $\bra{\vec\alpha}$ is the identity. 
 
 The proof is by induction. Let $u$ be the largest element according to $\prec$. For any $\ket{\phi}_I$, uniformly strong determinism implies that $$\bra{\vec \alpha}\bra P E_GN_{V\setminus I} \ket {\phi}_I \simeq  C_u\bra{\vec \alpha}\bra P \bar \lambda_u E_GN_{V\setminus I} \ket {\phi}_I$$ where $\bar \lambda_u$ is any Pauli operator which is not proportional to a Pauli operator in $\lambda_u$ (typically $\bar \lambda_u = Z$ when $\lambda_u=\{X,Y\})$. So $$\bra{\vec \alpha}\bra P E_GN_{V\setminus I} \ket {\phi}_I \simeq  \bra{\vec \alpha}C_u\bra P \bar \lambda_u E_GN_{V\setminus I} \ket {\phi}_I$$
 
By applying Lemma~\ref{lem:meas} for each non-Pauli measurement, $\exists L\subseteq
\{u~|~|\lambda_u|=2\}$, $\exists x\in\{0,1\}^{|L|}$, $\exists \ket{\psi}$ s.t.  $\bra P
E_GN_{V\setminus I}\ket {\phi}_I\simeq (B\ket x_{L}) \otimes \ket \psi$ and $C_u\bra P \bar
\lambda_u E_GN_{V\setminus I} \ket {\phi}_I\simeq (B\ket {\bar x}_{L})\otimes \ket \psi$, where
$\forall v\in L$,
$$~B_v:=\begin{cases} I&\text{if  $\lambda_v=\{X,Y\}$}\\  H&\text{if  $\lambda_v=\{Y,Z\}$}\\  \sqrt
X&\text{if  $\lambda_v=\{Z,X\}$ where for a Pauli $\sigma$, $\sqrt \sigma = \frac  {I + i \sigma}
2$} \end{cases}.$$
 
 So $\bra P E_GN_{V\setminus I}\ket {\phi}_I \simeq C_u\bar \lambda_u(B\ket {\bar x}_{L}\otimes \ket \psi)$
\begin{itemize}
	\item If $u\notin L$, $\bra{P}E_GN_{V\setminus I}\ket {\phi}_I \simeq B\ket {\bar x}_{L}\otimes C_u\bar \lambda_u\ket \psi$, so $B\ket {\bar x}_{L}\otimes C_u\bar \lambda_u\ket \psi \simeq B\ket x_{L} \otimes \ket \psi$ which implies $L=\emptyset$. 
	\item If $u\in L$, $\bra{P}E_GN_{V\setminus I}\ket {\phi}_I \simeq \bar \lambda_uB\ket {\bar
		x}_{L}\otimes C_u\ket \psi$, so $\bar \lambda_uB \ket {\bar x}_{L}\otimes C_u\ket \psi \simeq
		B\ket x_{L} \otimes \ket \psi$. Notice that $\bra x_L B^\dagger \bar \lambda_uB\ket {\bar x}_L =
		0$, indeed:
\subitem $\bullet$ if $\lambda_u = \{X,Y\}$ then $B_u =I$ and $\bar \lambda_u\simeq Z$ so $\bra x_u B_u^\dagger \bar \lambda_uB_u\ket {\bar x}_u=  0$.
\subitem $\bullet$ if $\lambda_u = \{Z,X\}$ then $B_u =\sqrt X$ and $\bar \lambda_u\simeq Y$ so $\bra x_u B_u^\dagger \bar \lambda_uB_u\ket {\bar x}_u=  0$ since $\sqrt X^\dagger Y \sqrt X\simeq Z$. 
\subitem $\bullet$ if $\lambda_u = \{Y,Z\}$ then $B_u =H$ and $\bar \lambda_u\simeq X$ so $\bra x_u B_u^\dagger \bar \lambda_uB_u\ket {\bar x}_u=  0$ since $H X H\simeq Z$. 
\end{itemize}
It implies $L=\emptyset$. 
As a consequence $\bra PE_GN_{V\setminus I}\ket {\phi}_I\simeq C_u\bra P \bar \lambda_u E_GN_{V\setminus I}\ket {\phi}_I$
 
 When $\ket {\phi}_I = \ket {+_I}$,  we have $\bra P\ket G\simeq C_u\bra P \bar \lambda_u \ket G$.

 \begin{itemize}
\item  If $|\lambda_u|=2$ then $ C_u \bar \lambda_u \bra P \ket G\simeq \bra P\ket G$, so according to Lemma~\ref{lem:meastab}\begin{equation}C_u\bar \lambda_u\simeq X_{S\setminus x(P)} Z_{\odd{S}\setminus z(P)}\label{eqPF}\end{equation} with \begin{equation}S\cap z(P) = \odd{S} \cap x(P).\label{eqPF2}\end{equation} 
By considering the case $\ket{\phi}_I=\ket 0_I$, it also implies $S\subseteq I^c$.  Moreover $C_u$
should not act on the already measured qubits so we recover in the following the Pauli Flow
condition by defining $p(u):=S$. Indeed assume w.l.o.g.\ that $\lambda_u=\{X,Y\}$, so $\bar
\lambda_u = Z$, and according to Eq.~\ref{eqPF}, $u\in \odd S \setminus S$.

Moreover, for any $v\prec u$, with $v\in S\cup \odd S$, 

\subitem $\bullet$ if $v\in S\setminus \odd S$, Eq.~\ref{eqPF} implies that $v\in x(P)$ and Eq.~\ref{eqPF2} that $v\notin z(P)$, so $\lambda_v=\{X\}$;
\subitem $\bullet$ if $v\in S\cap \odd S$, Eq.~\ref{eqPF} implies that $v\in x(P)$ and $v\in z(P)$, so $\lambda_v=\{Y\}$;
\subitem $\bullet$ if $v\in \odd S\setminus S$, Eq.~\ref{eqPF} implies that $v\in z(P)$ and Eq.~\ref{eqPF2} that $v\notin x(P)$, so $\lambda_v=\{Z\}$.

 \item If $|\lambda_u|=1$, assume $\lambda_u = \{X\}$. Let $A$ be the set of all Pauli measured qubits but $u$. $$\bra {P_A} \ket G = {\ket +_u\otimes\bra {P} \ket G + e^{i\theta}\ket -_u  \otimes \bra {P} \bar \lambda_u\ket G}$$
 
 Strongness  implies that there exists a stabilizer $M$ of $\bra {P_A} \ket G$ which anticommutes with $\lambda_u$. According to Lemma~\ref{lem:meastab}, $\exists S_0$ s.t.  $S_0\cap z(P_A) = \odd{S_0}\cap x(P_A)$ and $M=X_{S_0\setminus x(P_A)}Z_{\odd{S_0}\setminus z(P_A)} $, moreover $u\in \odd{S_0}$ since  $M$ and $\lambda_u = \{X\}$ anticommute. 
 
 Since $X_{S_0\setminus x(P_A)}Z_{\odd{S_0}\setminus z(P_A)} \bra {P_A} \ket G \simeq \bra {P_A} \ket G$, we have 
 \begin{eqnarray*}
 \bra {P} \bar \lambda_u\ket G&\simeq& \bra +_u \bar \lambda_u\bra{P_A} \ket G\\ 
 &\simeq& \bra +_u  Z_u \bra{P_A} \ket G\\ 
  &\simeq& \bra +_u  Z_uX_{S_0\setminus x(P_A)}Z_{\odd{S_0}\setminus z(P_A)} \bra{P_A} \ket G\\ 
      &\simeq& X_{S_0\setminus x(P_A)\setminus u}Z_{\odd{S_0}\setminus z(P_A)} Z_u \bra {P} \ket G
  \end{eqnarray*}
 So $\bra {P} \ket G\simeq C_u X_{S_0\setminus x(P_A)\setminus u}Z_{\odd{S_0}\setminus z(P_A)} Z_u\bra {P} \ket G $
 
As a consequence, according to Lemma~\ref{lem:meastab}, $\exists S_1$ s.t.\ $S_1\cap z(P) = \odd{S_1}\cap x(P)$ and $$C_uX_{S_0\setminus x(P_A)\setminus u}Z_{\odd{S_0}\setminus z(P_A)}Z_u\simeq X_{S_1\setminus x(P)\setminus u}Z_{\odd{S_1}\setminus z(P)}$$
 
So $C_u =  X_{S_0\Delta S_1\setminus x(P)}Z_{\odd{S_0\Delta S_1}\setminus z(P)}Z_u$ .
 
 Moreover, with $\ket{\phi}_I=\ket 0_I$, we have $S_0\Delta S_1 \subseteq I^c$. We recover the Pauli Flow condition by defining $p(u):= S_0\Delta S_1$.
 \end{itemize}
\end{proof}

\section{Proof of Proposition~\ref{Paulifirstpattern}}\label{appendixthm}

\Paulifirstpattern*

As $\rhd$ is defined by a sequence of $\rhd_u$, we 
just need to prove by induction  that if  $\mathcal{P}_0\rhd_u \mathcal{P}_1$ and $\mathcal{P}_0$ is robustly deterministic then so is $\mathcal{P}_1$ and $\interp{  \mathcal{P}_0}=  \interp {\mathcal{P}_1}$.

\begin{itemize}
\item Let  $\mathcal{P}_0=\X^{\s_v}_A\Z^{\s_v}_B\M_v^{\lambda, \theta_v}\mathcal P'_0$ and
	$\mathcal{P}_1=\X^{\s_v}_A\Z^{\s_v}_B\M_v^{\lambda, \theta_v}  \mathcal P'_1$  with  $\mathcal{P}'_0\rhd_u  \mathcal{P'}_1$. Since   $\mathcal P_0$ is robustly deterministic, so is  $\mathcal{P}'_0$, by induction  $\mathcal{P}'_1$ is robustly deterministic  and  $\interp{  \mathcal{P}'_0}=  \interp {\mathcal{P}'_1}$ hence 
$\interp{  \mathcal{P}_0}=  \interp {\mathcal{P}_1}$ and $\mathcal{P}_1$ is robustly deterministic.
\item Let  $\mathcal{P}_0= C^{s_u}\M_u^{P,\theta_u}\sigma^{s_v}_uR^{s_v}\M_v^{\{Q,Q'\}, \theta_v} \mathcal P$
and   $\mathcal{P}_1= 
	R'^{s_v}\M_v^{\{Q,Q'\}, \theta_v} C'^{s_u}\M_u^{P,\theta_u} \mathcal P$.
	
		Let $r$ and $c$ in $\{0, 1\}$ s.t. $R^2 = (-1)^rI$ and $C^2 = (-1)^cI$.
  Let $\ket \varphi$ be the state of the system after the application of $\mathcal P$.
  There exist four (not necessarily normalized) vectors $\ket{\phi_{i,j}}$ s.t. 
$$\begin{aligned}
\ket \varphi = &\ket{+^{Q,Q'}_0}_v\ket{+^P_0}_u \ket
{\phi_{00}}+\ket{+^{Q,Q'}_0}_v\ket{-^P_0}_u\ket {\phi_{01}}\\
&+\ket{-^{Q,Q'}_0}_v\ket{+^P_0}_u\ket {\phi_{10}}+\ket{-^{Q,Q'}_0}_v\ket{-^P_0}_u\ket {\phi_{11}}
\end{aligned}$$

Let $\ket {\Psi^{\theta_v}_{s_v}}$ be the state of the system after the application of the
  measurements of $v$, and appropriate corrections. When  $\theta_v=0$ (i.e. $v$ is
  $Q$-measured) and $s_v=0$,  $\ket {\Psi^0_0} =\ket{+^P_0}_u\ket {\phi_{00}} + \ket{-^P_0}_u\ket
  {\phi_{01}}$.

  When followed by the $P$-measurement of $u$ and the corresponding corrections, the two
  possible states, $\ket{\phi_{00}}\simeq C\ket{\phi_{01}}$ because of the strong
  determinism of $\mathcal P'$. Thus $\exists \alpha$ s.t. $\ket{\phi_{01}} =
  e^{i\alpha}C\ket{\phi_{00}}$ and $$\ket {\Psi^0_0} = \ket{+^P_0}_u\ket {\phi_{00}} + e^{i\alpha}\ket{-^P_0}_uC\ket {\phi_{00}}.$$

  Similarly, for $\theta_v=\pi$, we obtain that $\exists \beta$ s.t. $\ket{\phi_{11}} =
  e^{i\beta}C\ket{\phi_{10}}$. Thus,
$$\begin{aligned}
\ket \varphi =& \ket{+^{Q,Q'}_0}_v\ket{+^P_0}_u\ket {\phi_{00}}+e^{i\alpha}
\ket{+^{Q,Q'}_0}_v\ket{-^P_0}_uC\ket {\phi_{00}}\\
&+ \ket{-^{Q,Q'}_0}_v\ket{+^P_0}_u\ket {\phi_{10}}+e^{i\beta} \ket{-^{Q,Q'}_0}_v\ket{-^P_0}_uC\ket {\phi_{10}}
\end{aligned}$$

\subitem[i] Assume $\sigma P=- P \sigma$:
	
We consider the case $\theta_v=0$ and $s_v=1$ which, after the appropriate corrections, produces the state: 
  $$\ket {\Psi^0_1} \simeq \ket{-^P_0}_uR\ket{\phi_{10}} + (-1)^d e^{i\beta }\ket{+^P_0}_uRC\ket{\phi_{10}}$$
	with $d\in \{0,1\}$
\footnote{According to the definition of the measurement basis
		$\left\{\ket{+^P_{\theta_u}},\ket{-^P_{\theta_u}}\right\}$ (Equation~\ref{measurementBasis}), each
$d\in\{0,1\}$ corresponds to a choice of $\sigma$ anticommuting with $P$, more precisely we have
$d=0$ if and only if $\sigma=Z$ or ($\sigma=X$ and $P=Z$).}

Since $\mathcal P_0$ is robustly deterministic, the possible states of the register after the
measurement of $v$ (and the corresponding corrections) should be the same up to a global phase
($\ket {\Psi^0_1}\simeq \ket{\Psi^0_0}$). As a consequence, $\exists \gamma$ s.t.\
$(-1)^de^{i\beta}RC\ket{\phi_{10}}=e^{i\gamma}\ket{\phi_{00}}$ and $R\ket{\phi_{10}} =
e^{i(\alpha+\gamma)}C\ket {\phi_{00}}$,

So $R\ket{\phi_{10}} =(-1)^de^{i(\alpha+\beta)}CRC\ket {\phi_{10}}=
(-1)^{a+d}e^{i(\alpha+\beta)}C^2R\ket {\phi_{10}}=  (-1)^{a+c+d} e^{i(\alpha+\beta)}R\ket
{\phi_{10}}$, where $a\in \{0,1\}$ is s.t.  $RC=(-1)^{a} CR$.

It implies $\beta = -\alpha + (a+c+d)\pi \bmod 2\pi$,  $C\ket{\phi_{10}} =
(-1)^{d+r}e^{i(\gamma-\beta)}R\ket{\phi_{00}}$, and $\ket{\phi_{10}}=
(-1)^{r}e^{i(\alpha+\gamma)}RC\ket{\phi_{00}}$. 

As a consequence, 
\begin{eqnarray*}\ket \varphi &= & \ket{+^{Q,Q'}_0}_v\ket{+^P_0}_u\ket {\phi_{00}}
+e^{i\alpha} \ket{+^{Q,Q'}_0}_v\ket{-^P_0}_uC\ket {\phi_{00}}\\&&+
 (-1)^{r}e^{i(\alpha+\gamma)}\ket{-^{Q,Q'}_0}_v\ket{+^P_0}_uRC\ket {\phi_{00}}
 +(-1)^{r+d}e^{i\gamma}\ket{-^{Q,Q'}_0}_v\ket{-^P_0}_uR\ket {\phi_{00}}
 \end{eqnarray*}
So, with $\delta= \gamma+r\pi$, we get 
\begin{eqnarray*}\ket \varphi &= & \ket{+^{Q,Q'}_0}_v\ket{+^P_0}_u\ket {\phi_{00}}
+
 e^{i(\alpha+\delta)}\ket{-^{Q,Q'}_0}_v\ket{+^P_0}_uRC\ket {\phi_{00}}\\&&
+e^{i\alpha} \ket{+^{Q,Q'}_0}_v\ket{-^P_0}_uC\ket {\phi_{00}}
 +(-1)^de^{i\delta}\ket{-^{Q,Q'}_0}_v\ket{-^P_0}_uR\ket {\phi_{00}}
 \end{eqnarray*}

We consider two cases depending whether $\ket {\phi_{00}}$ and $RC\ket{\phi_{00}}$ are independent or not. 

\begin{itemize}
  \item If $\ket {\phi_{00}}$ and $RC\ket{\phi_{00}}$ are not colinear, then we consider
  the measurement of $v$ according to $\theta_v = \frac \pi 2$,
  
  Since for any $Q\neq Q'$,  $\bra{+_{\pi/2}^{Q,Q'}} \ket{+_0^{Q,Q'}}=\bra{-_{\pi/2}^{Q,Q'}} \ket{-_0^{Q,Q'}} = \frac{1-i}{2}$ and $\bra{+_{\pi/2}^{Q,Q'}} \ket{-_0^{Q,Q'}}=\bra{-_{\pi/2}^{Q,Q'}} \ket{+_0^{Q,Q'}} = \frac{1+i}{2}=\frac{1-i}{2}i$, 
  when the outcome is $s_v=0$,  the resulting state is,
$$\begin{aligned}
\ket{\Psi_0^{\frac \pi 2}} \simeq &\ket{+^P_0}_u\ket {\phi_{00}}+e^{i(\alpha+\delta')}\ket{+^P_0}_u
RC\ket{\phi_{00}}\\
&+e^{i\alpha}\ket{-^P_0}_uC\ket {\phi_{00}}+(-1)^de^{i\delta'}\ket{-^P_0}_uR\ket {\phi_{00}}
\end{aligned}$$
where $\delta'=\delta+\pi/2$.

 When followed by the $P$-measurement of $u$ and the corresponding corrections, the two possible states:

  $$\ket {\phi_{00}}+e^{i(\alpha+\delta')} RC\ket{\phi_{00}} \quad \text{and} \quad
    (-1)^ce^{i\alpha}\ket {\phi_{00}}+(-1)^{a+d}e^{i\delta'}RC\ket {\phi_{00}}$$
are equal up to a global phase because of the strong determinism of $\mathcal P_0$. Thus $\exists \epsilon$ s.t.

    $$e^{i\epsilon}(\ket {\phi_{00}}+e^{i(\alpha+\delta')} RC\ket{\phi_{00}})=
    (-1)^ce^{i\alpha}\ket {\phi_{00}}+(-1)^{a+d} e^{i\delta'}RC\ket {\phi_{00}}$$
It implies, since $\ket {\phi_{00}}$ and $RC\ket{\phi_{00}}$ are not colinear, that $\epsilon = \alpha +c\pi \bmod 2\pi$
    and $ \epsilon + \alpha + \delta' = (a+d)\pi +\delta' \bmod 2\pi$, so $2\alpha  =
    (a+c+d)\pi\bmod 2\pi$. 

As a consequence, 
$$\begin{aligned}
\ket \varphi = &\ket{+^{Q,Q'}_0}_v\ket{+^P_0}_u\ket {\phi_{00}} +
e^{i(\alpha+\delta)}\ket{-^{Q,Q'}_0}_v\ket{+^P_0}_uRC\ket {\phi_{00}}\\
&+(-1)^{a+c+d}e^{-i\alpha} \ket{+^{Q,Q'}_0}_v\ket{-^P_0}_uC\ket {\phi_{00}}\\
& +(-1)^de^{i\delta}\ket{-^{Q,Q'}_0}_v\ket{-^P_0}_uR\ket {\phi_{00}}
\end{aligned}$$

    Hence, the patterns $C^{s_u}M_u^{P,\theta_u} \mathcal P$ and  $\mathcal
    P_1=C^{s_v}R^{s_v}\M_v^{\{Q,Q'\}, \theta_v} C^{s_u}\M_u^{P,\theta_u} \mathcal P$ are strongly deterministic, so $\mathcal P_1$       is robustly deterministic.  

Indeed after the  measurement  of $u$ in $C^{s_u}M_u^{P,\theta_u} \mathcal P$ the two possible states after correction are
 $\ket{\varphi_0^0}=\ket{+^{Q,Q'}_0}_v\ket {\phi_{00}}+ e^{i(\alpha+\delta)}\ket{-^{Q,Q'}_0}_vRC\ket {\phi_{00}}$ and 
 $\ket{\varphi_0^1}=(-1)^{a+c+d}e^{-i\alpha} \ket{+^{Q,Q'}_0}_vC^2\ket {\phi_{00}}
 +(-1)^de^{i\delta}\ket{-^{Q,Q'}_0}_vCR\ket {\phi_{00}}=(-1)^{a+d}e^{-i\alpha}\ket{\varphi_0^0}$, as $RC=(-1)^{a} CR$.
 For the full pattern, we check that for any $v$ measurement  angle $\theta_v$, the branches (after both measurements) with same $s_u$ and $s_v$ in  $ \mathcal P_0$ and  $\mathcal P_1$ are equal up to a global phase.
 \begin{itemize}
 \item When $s_u=s_v=0$ both branches are obviously equal.
 \item When $s_v=0$, and $s_u=1$ the resulting states are respectively $C\bra {-_0^P}_u
 	\bra{+_{\theta_v}^{Q,Q'}}_v \ket \varphi$ for $\mathcal P_0$ and  $\bra{+_{\theta_v}^{Q,Q'}}_vC\bra {-_0^P}_u   \ket \varphi$ for $\mathcal P_1$ which are equal.
 \item When $s_v=1$ and $s_u=0$  the resulting states are respectively $\bra {+_0^P}_u
 	 \sigma_uR\bra{-_{\theta_v}^{Q,Q'}}_v \ket \varphi$ for $\mathcal P_0$ and
 	 $CR\bra{-_{\theta_v}^{Q,Q'}}_v\bra {+_0^P}_u   \ket \varphi$ for $\mathcal P_1$. The strong
 	 determinism of $C^{s_u}M_u^{P,\theta_u} \mathcal P$ implies that  $C\bra {-_0^P}_u   \ket \varphi \simeq \bra {+_0^P}_u   \ket \varphi$ so  $\bra {-_0^P}_u   \ket \varphi \simeq C\bra {+_0^P}_u   \ket \varphi$. Moreover $\bra {+_0^P}_u   \sigma_u\simeq \bra{-_0^P}_u$ since $\sigma$ and $P$ anticommute, as a consequence the two branches are equal up to a global phase.
  \item Similarly , when $s_u=s_v=1$ both branches are  equal up to a global phase.
  \end{itemize}
  Thus the strong determinism of $\mathcal P_0$ ensures the strong determinism of $\mathcal P_1$, moreover $\interp{\mathcal P_0}=\interp{\mathcal P_1}$.
\item  If $\ket {\phi_{00}}$ and $RC\ket{\phi_{00}}$ are colinear, then $\ket {\phi_{00}}$ is an eigenvector of $RC$, so $\exists k\in \{0,1,2,3\}$ s.t. $RC\ket{\phi_{00}} = i^k\ket {\phi_{00}}$, then $C\ket{\phi_{00}} = i^{2r+k}R\ket {\phi_{00}}$, and 
\begin{eqnarray*}\ket \varphi &= & \ket{+^{Q,Q'}_0}_v\ket{+^P_0}_u\ket {\phi_{00}}
+
 i^ke^{i(\alpha+\delta)}\ket{-^{Q,Q'}_0}_v\ket{+^P_0}_u\ket {\phi_{00}}\\&&
+ i^{2r+k}e^{i\alpha} \ket{+^{Q,Q'}_0}_v\ket{-^P_0}_uR\ket {\phi_{00}}
 +(-1)^de^{i\delta}\ket{-^{Q,Q'}_0}_v\ket{-^P_0}_uR\ket {\phi_{00}}\\
 &=& \left(\ket{+^{Q,Q'}_0}_v +i^ke^{i(\alpha+\delta)}\ket{-^{Q,Q'}_0}_v\right)\ket{+^P_0}_u\ket{\phi_{00}}\\&&+i^{2r+k}e^{i\alpha}\left(\ket{+^{Q,Q'}_0}_v +i^{2r-k+2d}e^{i(\delta-\alpha)}\ket{-^{Q,Q'}_0}_v\right)\ket{-^P_0}_uR\ket {\phi_{00}}
 \end{eqnarray*}

Strongness implies
$$\begin{aligned}
\left|\bra {+^{Q,Q'}_{\pi/2}} \left(\ket{+^{Q,Q'}_0}_v
+i^ke^{i(\alpha+\delta)}\ket{-^{Q,Q'}_0}_v\right)\right| = \\
\left|\bra {-^{Q,Q'}_{\pi/2}}
\left(\ket{+^{Q,Q'}_0}_v +i^ke^{i(\alpha+\delta)}\ket{-^{Q,Q'}_0}_v\right)\right|
\end{aligned}$$
and
$$\begin{aligned}
\left|\bra {+^{Q,Q'}_{\pi/2}} \left(\ket{+^{Q,Q'}_0}_v
+i^{2r-k+2d}e^{i(\delta-\alpha)}\ket{-^{Q,Q'}_0}_v\right)\right|=\\
\left|\bra {-^{Q,Q'}_{\pi/2}} \left(\ket{+^{Q,Q'}_0}_v +i^{2r-k+2d}e^{i(\delta-\alpha)}\ket{-^{Q,Q'}_0}_v\right)\right|,
\end{aligned}$$
so  
$\exists \ell, m \in \{0,1\}$ s.t. $i^ke^{i(\alpha+\delta)}= (-1)^\ell i$,  $e^{i(\delta-\alpha-r\pi-d\pi
-k\frac \pi 2 )} = (-1)^mi$, which implies $2\delta = (r+m+\ell+d+1)\pi \bmod 2\pi$, and $2\alpha =
(r+m+\ell+d+k)\pi \bmod 2\pi$.

 As a consequence,
$$\begin{aligned}
\ket \varphi =& \left(\ket{+^{Q,Q'}_0}_v +(-1)^\ell
i\ket{-^{Q,Q'}_0}_v\right)\ket{+^P_0}_u\ket{\phi_{00}}\\
&+ i^{2r+k}e^{i\alpha}\left(\ket{+^{Q,Q'}_0}_v +(-1)^m i\ket{-^{Q,Q'}_0}_v\right)\ket{-^P_0}_uR\ket {\phi_{00}}
\end{aligned}$$

Thus
$$\ket{\Psi_0^{\frac \pi 2}} =  {\sqrt 2} \left(e^{(-1)^\ell \frac \pi 4i}\ket{+^P_0}_u\ket{\phi_{00}}
+i^{2r+k}e^{i\alpha}    e^{(-1)^m \frac \pi 4i}   \ket{-^P_0}_uR\ket{\phi_{00}}\right)$$ 

and 
$$\ket{\Psi_1^{\frac \pi 2}} = {\sqrt 2} \left(e^{-(-1)^\ell \frac \pi 4i}\ket{-^P_0}_uR\ket{\phi_{00}}
+(-1)^di^{k}e^{i\alpha}    e^{-(-1)^m \frac \pi 4i}   \ket{+^P_0}_u\ket {\phi_{00}}\right)$$

Stepwise strong determinism implies that
$\exists \epsilon$ s.t. $e^{i\epsilon} e^{(-1)^\ell \frac \pi 4i}
=(-1)^di^{k}e^{i\alpha} e^{-(-1)^m \frac \pi 4i}$
and
$e^{i\epsilon} i^{2r+k}e^{i\alpha} e^{(-1)^m \frac \pi 4i} = e^{-(-1)^\ell \frac \pi 4i}$.
As a consequence $2\alpha = (r+k+d)\pi\bmod{2\pi}$.
Combining with previous equations on $2\alpha$, namely $2\alpha = (r+m+\ell+d+k)\pi \bmod 2\pi$, we get $m=\ell\bmod 2$. As a consequence: 
$$\begin{aligned}
\ket \varphi =& \left(\ket{+^{Q,Q'}_0}_v +(-1)^\ell i\ket{-^{Q,Q'}_0}_v\right)\otimes\left(\ket{+^P_0}_u\ket{\phi_{00}}+  i^{2r+k}e^{i\alpha}\ket{-^P_u}R\ket {\phi_{00}}\right)
\end{aligned}$$
    Hence, the pattern $C^{s_u}M_u^{P,\theta_u}\mathcal P$  is strongly deterministic as $CR\ket{\phi_{00}}\simeq \ket{\phi_{00}}$,
         and  the four branches $\mathcal P_1=C^{s_v}R^{s_v}\M_v^{\{Q,Q'\}, \theta_v}
         C^{s_u}\M_u^{P,\theta_u} \mathcal P$ are equal up to a global phase to the corresponding  branches 
         of $\mathcal P_0$ which guarantees  strong determinism of $\mathcal P_1$   and that $\interp{\mathcal P_1}=\interp{\mathcal P_0}$. Therefore $\mathcal P_1$  is robustly deterministic.
\end{itemize}

\subitem[ii] Otherwise $\sigma P=P\sigma$,

Now, we consider the case $\theta_v=0$ and $s_v=1$ which, after the appropriate
  corrections, produces the state:
  $$\ket {\Psi^0_1} \simeq \ket{+^P_0}_uR\ket{\phi_{10}} +  (-1)^{d'}e^{i\beta}\ket{-^P_0}_uRC\ket{\phi_{10}}$$
	with $d'\in \{0,1\}$ s.t. $d'=0$ iff $\sigma=I$.
	
Since $\mathcal{P}_0$ is stepwise and strong deterministic the possible states of the
  register after the measurement of $v$ (and the corresponding corrections) should be the
  same up to a global phase
$\left(\ket {\Psi^0_1}\simeq \ket{\Psi^0_0}\right)$.
  As a consequence, $\exists \gamma$ s.t. $R\ket{\phi_{10}}=e^{i\gamma}\ket{\phi_{00}}$ and
	$(-1)^{d'}e^{i\beta}RC\ket{\phi_{10}} = e^{i(\alpha+\gamma)}C\ket {\phi_{00}}$, so
	$e^{i(\beta+\gamma)}= (-1)^{a+d'}e^{i(\alpha+\gamma)}$ where $a\in \{0,1\}$ is s.t.  $RC=(-1)^{a}CR$. 
	It implies $\beta = \alpha + (a+d'){\pi} \bmod 2\pi$.

  As a consequence, 
$$\begin{aligned}
\ket \varphi = & \ket{+^{Q,Q'}_0}_v\ket{+^P_0}_u\ket {\phi_{00}} +e^{i\alpha} \ket{+^{Q,Q'}_0}_v\ket{-^P_0}_uC\ket {\phi_{00}}\\
&+ (-1)^{r}e^{i\gamma}\ket{-^{Q,Q'}_0}_v\ket{+^P_0}_uR\ket {\phi_{00}}\\
&+(-1)^{r}e^{i(\alpha +\gamma+(a+d'){\pi})}\ket{-^{Q,Q'}_0}_v\ket{-^P_0}_uCR\ket {\phi_{00}}
\end{aligned}$$
  So, with $\delta = \gamma+r\pi$ we get
  \begin{eqnarray*}\ket \varphi &= & \ket{+^{Q,Q'}_0}_v\ket{+^P_0}_u\ket {\phi_{00}}
+ e^{i\delta}\ket{-^{Q,Q'}_0}_v\ket{+^P_0}_uR\ket {\phi_{00}}
\\&&
+e^{i\alpha} \ket{+^{Q,Q'}_0}_v\ket{-^P_0}_uC\ket {\phi_{00}}
+e^{i(\alpha +\delta + (a+d'){\pi})}\ket{-^{Q,Q'}_0}_v\ket{-^P_0}_uCR\ket {\phi_{00}}
 \end{eqnarray*}
 
\subitem[ii.1]  If $a+d'=0$ we see that  the patterns $C^{s_u}M_u\mathcal P$ and  $\mathcal P_1=R^{s_v}\M_v^{\{Q,Q'\}, \theta} C^{s_u}\M_u^{P,\alpha} \mathcal P$ are strongly deterministic, so $\mathcal P_1$ 
      is robustly deterministic. 

Indeed after the  measurement  of $u$ in $C^{s_u}M_u^{P,\theta_u} \mathcal P$ the two possible states after correction are
 $\ket{\varphi_0^0}=\ket{+^{Q,Q'}_0}_v\ket {\phi_{00}}+ e^{i\delta}\ket{-^{Q,Q'}_0}_vR\ket {\phi_{00}}$ and 
 $\ket{\varphi_0^1}=e^{i\alpha} \ket{+^{Q,Q'}_0}_vC^2\ket {\phi_{00}}
 +e^{i(\alpha+\delta)}\ket{-^{Q,Q'}_0}_vC^2R\ket {\phi_{00}}=(-1)^{c}e^{i\alpha}\ket{\varphi_0^0}$, as $C^2=(-1)^{c} I$.
 For the full pattern, we check that for any $v$ measurement  angle $\theta_v$, the branches (after both measurements) with same $s_u$ and $s_v$ in  $ \mathcal P_0$ and  $\mathcal P_1$ are equal up to a global phase.
 \begin{itemize}
 \item When $s_u=s_v=0$ both branches are obviously equal.
 \item When $s_v=0$, and $s_u=1$ the resulting states are respectively $C\bra {-_0^P}_u
 	\bra{+_{\theta_v}^{Q,Q'}}_v \ket \varphi$ for $\mathcal P_0$ and  $\bra{+_{\theta_v}^{Q,Q'}}_vC\bra {-_0^P}_u   \ket \varphi$ for $\mathcal P_1$ which are equal.
 \item When $s_v=1$ and $s_u=0$  the resulting states are respectively $\bra {+_0^P}_u
 	 \sigma_uR\bra{-_{\theta_v}^{Q,Q'}}_v \ket \varphi$ for $\mathcal P_0$ and
 	 $R\bra{-_{\theta_v}^{Q,Q'}}_v\bra {+_0^P}_u   \ket \varphi$ for $\mathcal P_1$. Since $\sigma$ and $P$ commute, $\bra {+_0^P}_u   \sigma_u\simeq \bra{+_0^P}_u$  as a consequence the two branches are equal up to a global phase.
  \item Similarly , when $s_u=s_v=1$ both branches are  equal up to a global phase.
  \end{itemize}
  Thus the strong determinism of $\mathcal P_0$ ensures the strong determinism of $\mathcal P_1$, moreover $\interp{\mathcal P_0}=\interp{\mathcal P_1}$.
        
     \subitem[ii.2]  Otherwise, $a+d'=1$.

  By measuring $v$ with $\theta= \pi/2$ one gets:
  $$\ket{\Psi_0^{\frac \pi 2}}\simeq(I+ ie^{i\delta}R)\ket{+^P_0}_u\ket {\phi_{00}}+
  e^{i\alpha}C(I -i e^{i\delta}R)\ket{-^P_0}_u\ket {\phi_{00}}$$

  Then by measuring $u$ with $P$, there exists $\eta$ s.t.
  $$(I+ ie^{i\delta}R)\ket {\phi_{00}}= e^{i(\alpha+\eta +c \pi) }(I
  -ie^{i\delta}R)\ket {\phi_{00}}$$
 So $\ket {\phi_{00}}$ and $R\ket {\phi_{00}}$ are  colinear, hence  $\ket {\phi_{00}}$ is an eigenvector of $R$, and $\exists k\in \{0,1,2,3\}$
    s.t. $R\ket{\phi_{00}} = i^k\ket {\phi_{00}}$, and
$$\begin{aligned}
\ket \varphi =& \left( \ket{+^{Q,Q'}_0}_v+e^{i(\delta+k\pi/2)} \ket{-^{Q,Q'}_0}_v\right)\ket{+^P_0}_u \ket {\phi_{00}}\\
&+e^{i\alpha}\left(\ket{+^{Q,Q'}_0}_v-e^{i(\delta+k\pi/2)} \ket{-^{Q,Q'}_0}_v\right)\ket{-^P_0}_u C\ket{\phi_{00}}.
\end{aligned}$$
 
Strongness implies
$$\left|\bra {+^{Q,Q'}_{\pi/2}} \left(\ket{+^{Q,Q'}_0}_v
+i^ke^{i\delta}\ket{-^{Q,Q'}_0}_v\right)\right|
= \left|\bra {-^{Q,Q'}_{\pi/2}} \left(\ket{+^{Q,Q'}_0}_v
+i^ke^{i\delta}\ket{-^{Q,Q'}_0}_v\right)\right|$$
so $\exists \ell,  \in \{0,1\}$ s.t. $i^ke^{i\delta}=(-1)^\ell$

 As a consequence,
\begin{eqnarray*}\ket \varphi &=& \left(\ket{+^{Q,Q'}_0}_v +(-1)^\ell \ket{-^{Q,Q'}_0}_v\right)\ket{+^P_0}_u\ket{\phi_{00}}+\\
&&e^{i\alpha}\left(\ket{+^{Q,Q'}_0}_v -(-1)^\ell \ket{-^{Q,Q'}_0}_v\right)\ket{-^P_0}_uC\ket {\phi_{00}}
\end{eqnarray*}

Hence, for $~C'\in\{\widetilde{ Q_v}C,\widetilde {Q'_v}C\}$
 where $\widetilde K:= \begin{cases}\X\Z&\text{if $K= \Y$}\\ K &\text{otherwise}\end{cases}$, 
 the pattern $C'^{s_u}M_u\mathcal P$  is robustly  deterministic.
 
Moreover, for any  $R'\in\{R,\emptyset\}$, the four branches of the pattern  ${\mathcal
P_1}=R'^{s_v}M_v^{Q,Q'} C'^{s_u}M_u\mathcal P$ are equal up to a global phase to the corresponding
branches of $\mathcal P_0$ which guarantees  strong determinism of $\mathcal P_1$   and that
$\interp{\mathcal P_1}=\interp{\mathcal P_0}$. Therefore $\mathcal P_1$  is robustly deterministic.
\end{itemize}

$\hfill\Box$

\section{Proof of Proposition~\ref{thmCNExtPauli}}
\label{anxthm:push}
\thmpush*

\begin{proof}
Let $m_{\mathcal P}= \sum_{u\in O^c,\, |\lambda_u|=1} |\{v\in O^c~|~v<u \text{ and } |\lambda_v|=2\}|$ the total distance of Pauli measurements to the end of the computation.
We show by induction on $m_{\mathcal P}$ that $\mathcal P$ has a Shadow Pauli Flow $(p,<)$ that induces $\mathcal P$ and such that Condition~(1) is satisfied, 
where   \textbf {Condition~(1)} is: 
If $ w_1, w_2\in O^c, w_2\in \infl(p(w_1))$ 
{ and } $w_2<w_1$ then $|\lambda_{w_1}|=1 $ and after the measurement\footnote{and the associated correction} of $w_2$,  either: 

\begin{minipage}{\textwidth}
\begin{itemize}
\item $w_1$ is an isolated qubit in the state  $\ket {+_{\delta-\alpha_{w_2}}^{\lambda_{w_2}}}$ 

\item  the state is of the form $\lambda_{w_1}(\alpha_{w_2}) \ket \phi$ where $\ket \phi$ does not depend on $\alpha_{w_2}$.  
\end{itemize} \hfill
\end{minipage}
 
Where $\delta \in [0,2\pi)$ 
 is a constant that only depends on $w_1$ and $w_2$ (but not on the measurement angles $\alpha$).

 If $m_{\mathcal P}=0$ then all the Pauli measurements are at the end of the computation so according
 to Proposition~\ref{thm:PFrd}, since $\mathcal P$ is deterministic it has a Pauli Flow which can be seen has an
 Shadow Pauli Flow that trivially satisfies the induction hypothesis. 

Otherwise there exists $u\in O^c$ s.t. $|\lambda_u|=1$ and its previous measurement is on a qubit $v$ s.t. $|\lambda_v|= 2$.  We consider all the patterns $\mathcal P'_i$ s.t. $\mathcal P \rhd_u \mathcal P'_i$. According to Proposition~\ref{Paulifirstpattern} 
all $\mathcal P'_i$ are robustly deterministic, so by induction hypothesis they have a Shadow Pauli Flow $(p_i',<')$ satisfying Condition~(1) 
where $<'$ is obtained from $<$ by exchanging $u$ and $v$. We use the notation $\mathcal P'$ and $p'$ when there is a single $\mathcal P'_i$. 
Let $C$ be the corrector of $u$ in $\mathcal P$ and $R\sigma_u$ be the corrector of $v$ in $\mathcal P$ (where $R$ is not acting on $u$), and $\lambda_v=\{Q_0,Q_1\}$. 

\begin{itemize}
\item If $\lambda_u \sigma = -\sigma\lambda_u$, then there is a single $\mathcal P'$ s.t. $\mathcal P \rhd_u \mathcal P'$. Let $C'$ (resp $R'$) be the corrector of $u$ (resp. $v$)  in $\mathcal P'$. 

 Let $p: w\mapsto \begin{cases} p'(w)&\text{if $w\neq v$}\\ 
p'(v)\Delta p'(u)& \text{otherwise}\end{cases}$

We show that $(p,<)$ is a Shadow Pauli Flow that satisfies Condition~(1) and that $(p,<)$ induces $\mathcal P$. Notice that it is enough to show that $(p,<)$ satisfies the Shadow Pauli Flow conditions for $u$ and $v$ as it has been modified only on those two qubits. 
First notice that according to Definition~\ref{defpush} the corrector $C'=C$ consistent with $p'(u)$
in $\mathcal P'$ does not act on $v$ so $v\notin p'(u)\cup Odd(p'(u))$ so $v\notin \infl(p'(u))$ and
therefore $v\notin \infl(p(u))$ as $p(u)=p'(u)$. Moreover, $C$ is consistent with $p(u)$ and
Condition~(1) is satisfied for $w_1=u$.

Regarding $v$, $R'=CR$ so $R=CR'$, as a consequence $p(v)$ induces $R$ as $p'(u)$ induces $C$ and $p'(v)$ induces $R'$. Therefore $(p,<)$ induces $\mathcal P$. Moreover, regarding Condition~(1) in $\mathcal P$ for $w_1=v$, 
we prove by contradiction that $w_2\notin \infl(p(v))$ 
  for any $w_2<v$.

Notice that if there exists $w_2<' u$ s.t. 
$w_2\in \infl(p'(u))$
 then according to Condition~(1) in $\mathcal P'$, either: 
\begin{itemize}
\item[$(i)$]
 The qubit $u$ is an isolated qubit in the state  $\ket {+_{\delta-\alpha_{w_2}}^{\lambda_{w_2}}}$
 (before the measurements of both $u$ and $v$). 
 Thus in $\mathcal P$, the state of $u$ after the measurement of $v$ does not depend on the outcome of the measurement $v$. Thus by determinism the state of $u$ must be an eigenvector  of $\sigma_u$, therefore $\sigma_u \notin \lambda_{w_2}$ and $\delta-\alpha_{w_2}=0 \mod \pi$ which contradicts the uniform determinism.  
  So there is no $w_2<' u$ s.t.
 $w_2\in \infl(p'(u))$,  
   hence no $w_2< v$ s.t. 
 $w_2\in \infl(p(v))$, 
  thus Condition~(1) is also satisfied for $w_1=v$. 
\item [$(ii)$] Or,  the state after the  measurement of $w_2$  is of the form $\lambda_{u}(\alpha_{w_2}) \ket \phi$ where $\ket \phi$ does not depend on $\alpha_{w_2}$. Thus, in $\mathcal P$, the measurement of $v$ after corrections produces either $\lambda_{u}(\alpha_{w_2})  \bra {+_{\alpha_v}^{\lambda_v}}_v\ket \phi$ or $R\sigma_u\lambda_{u}(\alpha_{w_2})  \bra {-_{\alpha_v}^{\lambda_v}}_v\ket \phi =  e^{i\theta_0} \lambda_{u}(-\alpha_{w_2}) R\sigma_u \bra {-_{\alpha_v}^{\lambda_v}}_v\ket \phi$ that are equal up to a global phase. When $\alpha_{w_2}=0$, we get $\bra {+_{\alpha_v}^{\lambda_v}}_v\ket \phi  = e^{i\theta_1}\lambda_u(0)R\sigma_u \bra {-_{\alpha_v}^{\lambda_v}}_v\ket \phi$, thus
$\lambda_{u}(\alpha_{w_2}) \lambda_{u}(0) \bra {+_{\alpha_v}^{\lambda_v}}_v\ket \phi = e^{i\theta} \lambda_{u}(-\alpha_{w_2}) \bra {+_{\alpha_v}^{\lambda_v}}_v\ket \phi $ so $ \bra {+_{\alpha_v}^{\lambda_v}}_v\ket \phi $ is eigenvector of $ \lambda_{u}(2\alpha_{w_2})$. Therefore $u$ is in the state $\ket{+_{k\pi}^{\lambda_u}}$.
Hence the original pattern is not robust deterministic as $u$ is measured with Pauli $\lambda_u$.

\end{itemize}

\item Otherwise, if $RC=CR$ xor $\sigma=P$, 
 then there is a single $\mathcal P'$ s.t. $\mathcal P \rhd_u \mathcal P'$. Let $C'$ (resp $R'$) be the corrector of $u$ (resp. $v$)  in $\mathcal P'$. 
 
 We show that $(p',<)$ is a Shadow Pauli Flow that satisfies Condition~(1) and that $(p',<)$ induces $\mathcal P$. 
Similarly to the previous case, $C$ is consistent with $p'(u)$ and Condition~(1) is satisfied for
$w_1=u$. Moreover, $p'(v)$ also induces $R$ and Condition~(1) is also satisfied for $w_1=v$.

\item Finally, for the remaining case, there are several $\mathcal P'_i$~s.t.~$\mathcal P \rhd_u \mathcal P_i'$, Let $R'_0=R'_1=R$, $R'_2=R'_3=\emptyset$, $C'_0 = C'_2 = \widetilde {Q_v}C$ and $C'_1 = C'_3 = \widetilde{ Q'_v}C$ be respectively the correctors of $v$ and $u$ in the corresponding $\mathcal P_i'$. 

We show that $(p_0',<)$ is a Shadow Pauli Flow that satisfies Condition~(1)  and that $(p_0',<)$ induces $\mathcal P$. 
First notice that according to Definition~\ref{defpush} the corrector $C_0'=\widetilde {Q_v} C$, consistent with $p_0'(u)$ in $\mathcal P'_0$,  acts on $v$ so $v\in  p_0'(u)\cup \odd{p_0'(u)}$  so 
$v\in \infl(p_0'(u))$.

To show that $(p_0',<)$ is a Shadow Pauli Flow, it is enough to show that $D_v:= p'_2(v)$ satisfies
the Shadow Pauli Flow condition. Indeed in $\mathcal P'_2$, there is no Pauli correction for $v$,
thus all the other vertices in $D_v\cup \odd{D_v}$ are smaller than $v$ and cannot be anachronistically impacted by other corrections as they are Pauli-measured.

 To show that Condition~(1) is satisfied, 
w.l.o.g., assume $ Q_v=X$, $  Q'_v=Y$ and $P=Z$. So before measurement of $v$ in $\mathcal P_2'$,
$v$ is in a state in $\{\ket 0, \ket 1\}$. 

Let $\ket \varphi$ be the state of the system  before the measurement of $u$ in $\mathcal P'_2$.  
 There exist four (not necessarily normalized) vectors $\ket{\phi_{i,j}}$ s.t.
  $$\ket \varphi = \ket{0_v0_u}\ket {\phi_{00}}+\ket{0_v1_u}\ket {\phi_{01}}+\ket{1_v0_u}\ket {\phi_{10}}+\ket{1_v1_u}\ket {\phi_{11}}$$

 Let $\ket {\Psi_{s_u}}$ be the state of the system after the application of the
  measurements of $u$, and appropriate corrections.

  $\ket {\Psi_0} = \ket {0_v}\ket {\phi_{00}} + \ket {1_v}\ket {\phi_{10}}$. 

  $\ket {\Psi_1} = \tilde Q_v \ket {0_v}C\ket {\phi_{01}} +\tilde Q_v  \ket {1_v}C\ket {\phi_{11}}=\ket {1_v}C\ket {\phi_{01}} + \ket {0_v}C\ket {\phi_{11}} $. 
 
	$\exists \beta$ s.t.\ $e^{i\beta} C\ket{\phi_{01}} =  \ket{\phi_{10}}$ and  $e^{i\beta} C
	\ket{\phi_{11}} =  \ket{\phi_{00}}$.

  Therefore,
  $$\ket \varphi = \ket{0_v0_u}\ket {\phi_{00}}+e^{-i\beta}\ket{1_v1_u} C\ket
  {\phi_{00}}+\ket{1_v0_u}\ket {\phi_{10}}+e^{-i\beta}\ket{0_v1_u} C\ket {\phi_{10}}$$
  
  with $\ket {\phi_{00}}=0$ or $\ket {\phi_{10}}=0$. w.l.o.g.\ we can assume that $\ket {\phi_{10}}=0$.
   
  $\ket \varphi = \ket{0_v0_u} \ket {\phi_{00}}+e^{-i\beta}\ket{1_v1_u} C\ket {\phi_{00}}$.  
  
We have   $\ket {\Psi_0} = \ket {0_v}\ket {\phi_{00}} $. As $\mathcal P_0'$  is robustly  deterministic:

    \begin{itemize}
    \item If $s_v=0$: $\bra{+_{\alpha_v}^{X,Y}} \ket   {\Psi_0} = \frac 1 {\sqrt 2}  \ket
			{\phi_{00}}$ and,
      \item If $s_v=1$: $R \bra{-_{\alpha_v}^{X,Y}} \ket  {\Psi_0} = \frac 1 {\sqrt 2} e^{-i \alpha_v} R  \ket {\phi_{00}}$.
  \end{itemize}
  
  Therefore robust determinism of  $\mathcal P_0'$ implies that      $R  \ket {\phi_{00}}=e^{i \gamma}   \ket {\phi_{00}}$. Notice that $\gamma$ does not depend on the particular angle of measurement $\alpha_v$.
  \begin{itemize}
    \item If  $\sigma=P$ and $RC=CR$,   $C$ and $R$ have same eigenvectors and   $C\ket {\phi_{00}}=e^{i \delta}   \ket {\phi_{00}}$. 
    Therefore $\ket \varphi = (\ket{0_v0_u} +e^{-i(\beta+\delta)}\ket{1_v1_u})\otimes  \ket {\phi_{00}}$. 
    After the measurement of $v$ in $\mathcal P$, we get 
   \begin{itemize}
   \item If $s_v=0$: $\bra{+_{\alpha_v}^{X,Y}} \ket   {\varphi}= ( \bra{+_{\alpha_v}^{X,Y}}0\rangle
   	 \ket {0_u} + e^{-i(\beta+\delta)} \bra{+_{\alpha_v}^{X,Y}}1\rangle \ket{1_u}) \otimes \ket
   	 {\phi_{00}} = \frac 1{\sqrt 2}(\ket {0_u} + e^{-i(\beta+\delta+\alpha_v)}  \ket{1_u}) \otimes
		 \ket {\phi_{00}}$.
      \item If $s_v=1$: $\sigma_uR\bra{-_{\alpha_v}^{X,Y}} \ket   {\varphi}= e^{i \gamma}(
      	\bra{-_{\alpha_v}^{X,Y}}0\rangle \ket {0_u} - e^{-i(\beta+\delta)}
      \bra{-_{\alpha_v}^{X,Y}}1\rangle \ket{1_u}) \otimes \ket {\phi_{00}} = \frac {e^{i
      \gamma}}{\sqrt 2}(\ket {0_u} + e^{-i(\beta+\delta+\alpha_v)}  \ket{1_u}) \otimes \ket
			{\phi_{00}}$.
      
      \end{itemize}
   So $u$ is separable and is in the state $\ket{+^{X,Y}_{\delta'-\alpha_{v}}}$.
    \item If $\sigma=I$ and $RC=-CR$,

   in ${\mathcal P}$ measuring $v$ first we get:
\begin{itemize} 
  \item If $s_v=0$: $\bra{+_{\alpha_v}^{X,Y}} \ket \varphi= \frac 1 {\sqrt 2} (\ket {0_u}   \ket
  {\phi_{00}}+e^{-i(\beta+\alpha_v)}\ket{1_u} C\ket {\phi_{00}})= Z_u(\alpha_v) ( \frac 1 {\sqrt 2}
  (\ket {0_u}   \ket {\phi_{00}}+e^{-i\beta}\ket{1_u} C\ket {\phi_{00}}))$ which satisfies the
  condition.
    \item If $s_v=1$: $R \bra{-_{\alpha_v}^{X,Y}} \ket \varphi=  \frac 1 {\sqrt 2} R(\ket{0_u}\ket {\phi_{00}} -e^{-i(\beta+\alpha_v)}\ket{1_u} C\ket {\phi_{00}})=  \frac {e^{i\gamma}} {\sqrt 2} (\ket{0_u}\ket {\phi_{00}} +e^{-i(\beta+\alpha_v)}\ket{1_u} C\ket {\phi_{00}})=  e^{i\gamma}Z_u(\alpha_v) ( \frac 1 {\sqrt 2} (\ket {0_u}   \ket {\phi_{00}}+e^{-i\beta}\ket{1_u} C\ket {\phi_{00}}))$ which satisfies the condition.
\end{itemize}

\end{itemize}  

\end{itemize}

\end{proof}

\end{document}